%% file: OscStrengthArxiv.tex
\documentclass[aps,pra,twocolumn,groupedaddress,nofootinbib]{revtex4}

\usepackage{amsmath,mathrsfs,graphics,epsfig,ulem,longtable,graphicx,epstopdf}
\usepackage[usenames]{color}

\newcommand{\bd}[1]{\mathbf{#1}}

\newcommand{\partd}[2]{\frac{\partial{#1}}{\partial{#2}}}
\newcommand{\fulld}[2]{\frac{d{#1}}{d{#2}}}
\newcommand{\fulldd}[2]{\frac{d^2{#1}}{d{#2}^2}}

\newcommand{\pd}[1]{\frac{\partial}{\partial{#1}}}
\newcommand{\pdd}[1]{\frac{\partial^2}{\partial{#1}^2}}
\newcommand{\pde}[2]{\frac{\partial^2}{\partial{#1}\partial{#2}}}

\newcommand{\gencoord}{{\rm X}}

\begin{document}

\title{Pseudospectral Calculation of Helium Wave Functions, Expectation Values, and Oscillator Strength}

\author{Paul E. Grabowski}
\email[]{grabowski@lanl.gov}
\affiliation{Department of Physics, Cornell University, Ithaca, NY 14853, USA}
\altaffiliation{Current Address: Computational Physics Group (CCS-2), Los Alamos National Laboratory, Mail Stop D413,  Los Alamos, NM 87545, USA}
\author{David F. Chernoff}
\email[]{chernoff@astro.cornell.edu}
\affiliation{Department of Astronomy, Cornell University, Ithaca, NY 14853, USA}

\date{\today}

\begin{abstract}

The pseudospectral method is a powerful tool for finding highly precise solutions of
Schr\"{o}dinger's equation for few-electron problems.
Previously we developed the method to calculate fully correlated
S-state wave functions for two-electron atoms
\cite{GrabowskiChernoff2010}. Here we extend the method's scope to
wave functions with non-zero angular momentum and test it on
several challenging problems.
One group of tests involves the
determination of the nonrelativistic electric dipole oscillator strength for the helium
$1^1$S $\to 2^1$P transition. The result achieved,
$0.27616499(27)$, is comparable to the best in the
literature. The formally equivalent length, velocity,
and acceleration expressions for the oscillator
strength all yield roughly the same accuracy because the numerical
method constrains the wave function errors in a local fashion.

Another group of test applications is comprised of well-studied leading
order finite nuclear mass and relativistic corrections
for the helium ground state.  A straightforward computation
reaches near state-of-the-art accuracy without
requiring the implementation of any special-purpose
numerics.

All the relevant quantities tested in this paper --
energy eigenvalues, S-state expectation values and
bound-bound dipole transitions for S and P states --
converge exponentially with increasing resolution and do so
at roughly the same rate. Each individual calculation
samples and weights the configuration space wave function
uniquely but all behave in a qualitatively similar
manner.  Quantum
mechanical matrix elements are directly and reliably
calculable with pseudospectral methods.

The technical discussion includes a prescription for
choosing coordinates and subdomains to achieve
exponential convergence when two-particle Coulomb
singularities are present. The prescription does not
account for the wave function's non-analytic behavior
near the three-particle coalescence which should eventually
hinder the rate of the convergence.  Nonetheless the
effect is small in the sense that ignoring the
higher-order coalescence does not appear to affect adversely
the accuracy of any of the quantities reported nor the
rate at which errors diminish.

\end{abstract}


\maketitle

\section{Introduction}

The aim of this work is to test and validate
the pseudospectral method as a high-precision
few-electron problem solver, capable of calculating state-of-the-art
precision matrix elements. 
The helium atom has been
studied extensively since the birth of quantum
mechanics and so makes a great testbed problem. 
High-precision work continues to this day
to infer fundamental constants such as the
fine structure constant (see Ref. \cite{PachuckiYerokhin2011}) and the electron-proton mass
ratio (see Ref. \cite{KorobovZhong2009}) by comparing theoretical and experimental measurements. 
Any theoretical method which may be applied to a variety of 
problems ({\it e.g.}  high-precision relativistic corrections, 
different interaction potentials,
excitation levels, symmetries, etc.) without tinkering
with or modifying the basis and which has direct,
rigorous control of local errors serves as a complementary
approach to the variational method.

Methods based on the variational principle, in which the
expectation value of the Hamiltonian is minimized with
respect to the parameters of a trial wave function, are 
the most widely used techniques for finding
an approximate representation of the
ground state.
The calculated energy is an upper bound to the exact
energy.\footnote{The method is not limited to ground
  states. A trial wave function, exactly orthonormal to
  all lower energy states, has calculated energy which
  is an upper bound to the exact result for the excited
  state.}
If one regards the best
approximate wave function as
first order accurate then the variationally determined
energy eigenvalue is second order accurate.  
Small errors in the
energy eigenvalue of a given state imply that the
square of the wave function is accurate
in the energy-weighted norm but it does not follow that
local wave function errors are also small. In practical
terms, while the variational
approach excels at determining energy eigenvalues it
does not generally achieve comparable accuracy
in quantum mechanical matrix elements formed
from the wave function.

To achieve ever-more accurate energies and/or wave
functions in the variational approach one must select a
sequence of trial functions capable of representing the
exact solution ever-more closely.  The choice of a good
sequence entails more than a little art and intuition,
especially for a nonstandard problem where one may have
only a vague idea what the ultimate limit looks
like. A sequence of increasing basis size $n$ may be said
to converge exponentially if the errors are
proportional to $e^{-a n}$ for some positive constant $a$. This
most favorable outcome is achieved only if the basis
can reproduce the analytic properties of the exact wave
function. Otherwise, convergence is expected to be
algebraic, i.e.  $\propto n^{-2}$, or worse.

Recently, we applied pseudospectral methods to solve
the nonrelativistic Schr\"{o}dinger equation for helium and the
negatively charged hydrogen ion with zero total angular momentum
\cite{GrabowskiChernoff2010}.  We found exponentially fast convergence
of most quantities of interest including the energy eigenvalues, local
energy errors ({\it e.g.} $(\hat{H} \Psi)/\Psi - E$ as a function of position) 
and Cauchy wave function differences. Only the error in
the logarithmic derivative near the triple coalescence point had
discernibly
slower convergence, presumably due to the logarithmic contributions
located there
\cite{Bartlett1937,Fock1954,Fock1958}.  
The key virtues of the pseudospectral approach were: no explicit
assumptions had to be made about the asymptotic behavior of the wave
function near cusps or at large distances, the Schr\"{o}dinger
equation was satisfied at all grid points, local errors decreased
exponentially fast with increasing resolution, and no fine tuning
was required.

In this article, we extend our previous work to higher angular
momentum calculations and utilize the results to evaluate matrix
elements for combinations of states. To be systematic, we consider two sorts
of matrix elements: the dipole absorption oscillator strength (between
S and P states) and first-order mass polarization and $\alpha^2$
relativistic corrections to the nonrelativistic finite-nuclear-mass
Hamiltonian (for the S ground state).  All have been the subject of
extensive
investigation.  Our main focus is on testing the pseudospectral method's
capabilities by recalculating these quantities and comparing to
effectively ``exact'' published results.

The plan of the paper is as follows. The first four sections are largely
background:
\S\ref{RevPSMethods} provides an overview of the pseudospectral method;
\S\ref{NR2electronatom} describes the two-electron atom, the Bhatia-Temkin
coordinate system, the expansion of the wave function in terms
of eigenstates and the form of the Hamiltonian;
\S\ref{DipoleReview} defines length, velocity and acceleration
forms for the oscillator strength and related sum rules.
The next two sections detail our pseudospectral
method of calculation and those readers primarily interested in seeing the
results may skip to \S\ref{EandOscResults}.
\S\ref{VarsAndDomains} gives a prescription for how to choose coordinates 
and subdomains for second order partial differential equations and
outlines the special coordinate choices
needed to deal with the Coulomb singularities.
\S\ref{BoundaryConditions} schematically describes how overlapping
and touching grids are coupled together and how
symmetry is imposed on the wave function.
\S\ref{EandOscResults} presents the first group of
test results on energies and oscillator strengths. The
convergence rate of all quantities is studied in detail.
\S\ref{CorrectionsHamiltonian} and \S\ref{MassPolRelCorrCalcs}
review lowest-order corrections to the
Hamiltonian due to finite nuclear mass and finite $\alpha$.
\S\ref{ExpectationValues} presents the second group of
test results for individual corrections
to the ground state of He.
\S\ref{Conclusions} summarizes the capabilities and promise
of the pseudospectral method.

The appendix is divided into four parts.
Appendix \ref{BTAppendix} gives the explicit form of the Hamiltonian operator used in this article.
Appendix \ref{MatrixMethods} describes how the Hamiltonian matrix problem is solved,
gives details of the eigenvalue solver method,
and how quantum mechanical matrix elements are calculated once the
wave function is determined.
Appendix \ref{BTOscStrength} gives the particular equations for calculating the oscillator strengths and expectation values.
Appendix \ref{OscStrengthTable} discusses and tabulates past work done to calculate oscillator strengths.

\section{Review of pseudospectral methods}
\label{RevPSMethods}

Pseudospectral methods have proven success in solving
systems of partial differential equations germane to the physics
in a wide variety of fields including fluid
dynamics \cite{CanutoEtAl1988}, general relativity
\cite{KidderFinn2000,PfeifferEtAl2003}, and quantum chemistry
\cite{Friesner1985,Friesner1986,Friesner1987,RingnaldaEtAl1990,GreeleyEtAl1994,MurphyEtAl1995,MurphyEtAl2000,KoEtAl2008,HeylThirumalai2009}.
Some problems in one-electron quantum mechanics 
\cite{Borisov2001,BoydEtAl2003} have been treated but only recently has
the method been applied to the case of fully correlated, multi-electron
atoms \cite{GrabowskiChernoff2010}.  Pseudospectral
methods are discussed in some generality in Refs.
\cite{Boyd2000,Fornberg1996,Orszag1980,PfeifferEtAl2003,NumericalRecipes,GrabowskiChernoff2010}.

The pseudospectral method is a grid-based finite difference method in
which the order of the finite differencing is equal to the resolution
of the grid in each direction.  As the grid size increases it becomes
more accurate than any fixed-order finite difference method. If a
solution is smooth over an entire domain (or smooth in each
subdomain) the pseudospectral method converges exponentially fast
to the solution. A spectral basis expansion
and a pseudospectral expansion of the same order 
are nearly equivalent having differences
that are exponentially small.

The grid points in the pseudospectral method are located at the roots
of Jacobi polynomials or their antinodes plus endpoints. They are
clustered more closely near the boundary of a domain than in its
center. Such an arrangement is essential for the method to limit
numerical oscillations sourced by singularities beyond the numerical
domain \cite{Fornberg1996sec34}. These singularities
typically occur in the analytic continuation of solutions to
non-physical regimes and/or from the extension of coordinates beyond the
patches on which they are defined to be smooth and differentiable. The grid point
arrangement facilitates a convergent representation of a function and
its derivative across the domain of interest. The interpolated function is
more uniformly accurate than is possible using an equal number of
equidistant points, as is typical for finite difference methods. 

Consider the problem of the pseudospectral representation of an operator 
like the Hamiltonian. The full domain is multi-dimensional but
focus for the moment on a single dimension
of the domain. Let $\{\bd{\gencoord}^k\}_{k=1,2,\ldots N}$ be
the roots of an $N$th order Jacobi polynomial enumerated by $k$. Let
$\bd{\gencoord}$ stand for an arbitrary coordinate value in the dimension
of interest. Define the one dimensional cardinal functions
\begin{equation}
C_j[\gencoord]=\prod_{\substack{k=1\\k\ne j}}^N \frac{\gencoord-\gencoord^k}{\gencoord^j-\gencoord^k}
\end{equation}
and note the relation
\begin{equation}
C_j[\gencoord^k]=\delta_j^k 
\end{equation}
follows. Now let the $n_d$-dimensional grid be the tensor product of
the individual, one dimensional coordinate grids labeled by $X_{(i)}$ for $i=1$ to $n_d$. 
The corresponding cardinal functions are
\begin{equation}
\label{EffectiveBasis}
{\cal C}_{J}[\bd{\gencoord}]=\prod_{i=1}^{n_d}C_{j_{(i)}}[\gencoord_{(i)}],
\end{equation}
where subscript $J=\{j_{(1)}, j_{(2)},\ldots,j_{(n_d)}\}$ and
unadorned $\bd{\gencoord}=\{\gencoord_{(1)},\gencoord_{(2)},\ldots,\gencoord_{(n_d)}\}$.
These multi-dimensional Cardinal functions have the property
\begin{equation}
{\cal C}_J[\bd{\gencoord}^K]=\delta_J^K,
\end{equation}
where the grid point $\bd{\gencoord}^K = \{\gencoord_{(1)}^{k_1},\gencoord_{(2)}^{k_2},\ldots,\gencoord_{(n_d)}^{k_{n_d}}\}$.
They
form a basis in the sense that a general function $f$ can be written
\begin{equation}
f[\bd{\gencoord}]=\sum_Jf[\bd{\gencoord}^J]{\cal C}_J[\bd{\gencoord}],
\end{equation}
where 
$f[{\gencoord}^J]$ is a pseudospectral coefficient (``pseudo'' because it is more easily identified as 
the function value at the grid point). 

Let the position $\bd{\gencoord}^K$ and cardinal ${\cal C}_J$ eigenstates
be denoted $|\bd{\gencoord}^K\rangle$ and $|\bd{\cal C}_J\rangle$, respectively.
The pseudospectral approximation to the Hamiltonian is
\begin{equation}
\label{PSMatrix}
\hat{H}_{PS}=\sum_{JK}|\bd{\gencoord}^K\rangle\langle\bd{\gencoord}^K|\hat{H}|{\cal C}_J\rangle\langle{\cal C}_J|,
\end{equation}
where $\hat{H}$ is the full Hamiltonian operator.
In practice, the matrix $\langle\bd{\gencoord}^K|\hat{H}_{PS}|{\cal C}_J\rangle$ is truncated and
then diagonalized to find the energy
eigenvalues.  When the wave function is represented by
a pseudospectral expansion the eigenvectors are simply the
function values at the grid points. In a spectral
representation, by contrast, the eigenvectors are sums of basis
functions. It is often more convenient and efficient
to work with the local wave function values directly.
On the other hand, the truncated operator
$\hat{H}_{PS}$ need not be Hermitian at finite resolution, a property
that may introduce non-physical effects, e.g. 
$\langle\bd{\gencoord}^K|\hat{H}_{PS}|{\cal C}_J\rangle$ may possess
complex eigenvalues. Generally, unphysical artifacts quickly reveal themselves
as resolution increases. An examination of the
eigenvalue spectrum shows that the complex
eigenvalues do not converge, permitting separation of
physical and unphysical values.

\section{The nonrelativistic two-electron atom}
\label{NR2electronatom}

Two-electron atoms are
three-particle systems requiring nine spatial coordinates for a full description. In the absence of external forces,
three coordinates are eliminated by taking out the center-of-mass motion.
In the infinite-nuclear-mass and nonrelativistic approximations the Hamiltonian is
\begin{equation}
\label{Hamiltonian0}
\hat{H}_0=-\frac{1}{2}(p_1^2+p_2^2)+\hat{V},
\end{equation}
where $\bd{p}_{1,2}$ are the momenta of the two electrons and the potential
is
\begin{equation}
\hat{V}=-\frac{Z}{r_1}-\frac{Z}{r_2}+\frac{1}{r_{12}},
\end{equation}
where $Z$ is the nuclear charge, and $r_1$,
$r_2$, and $r_{12}$ are the magnitudes of the vectors pointing from the
nucleus to each electron and of the vector pointing from one electron to
the other, respectively.
Here and throughout this article, atomic units are used. 
For the infinite-nuclear-mass approximation, the electron mass is
set to unity; for a finite nuclear mass,
the reduced mass of the electron and nucleus is set to one.
The fully correlated wave functions are six-dimensional at this
stage.

A further reduction is straightforward for S states.
Hylleraas \cite{Hylleraas1929} proposed the ansatz that the wave
function be written in terms of three
internal coordinates. Typical choices for these coordinates are $r_1$,
$r_2$, and $r_{12}$. Alternatively, $r_{12}$ may be replaced by
$\theta_{12}$, the angle between the two electrons. The S state
is independent of the remaining three coordinates that
describe the orientation of the triangle with vertices at the two
electrons and nucleus.

The situation for states of general angular momentum is more complicated.
Bhatia and Temkin \cite{BhatiaTemkin1964} introduced a particular set of
Euler angles $\{\Theta,\Phi,\Psi\}$ to describe the triangle's
orientation. 
They defined\footnote{The symbols used here are
  slightly different than those of \cite{BhatiaTemkin1964}
so that the equations can be written in a
  simplified form.}
a set of generalized spherical harmonics $D_{\kappa lm}^\nu$ which 
are eigenstates of operators for the total angular momentum, its $z$ component, total parity 
($\{\bd{r}_1,\bd{r}_2\}\to\{-\bd{r}_1,-\bd{r}_2\}$), and exchange ($\bd{r}_1\leftrightarrow \bd{r}_2$):
\begin{eqnarray}
\label{DEigenStates}
\hat{L}^2D_{\kappa lm}^\nu &=& l(l+1)D_{\kappa lm}^\nu\\
\hat{L}_zD_{\kappa lm}^\nu &=& mD_{\kappa lm}^\nu\\
\hat{\Pi}D_{\kappa lm}^\nu &=&(-1)^\kappa D_{\kappa lm}^\nu\\
\hat{{\cal E}}_{12}D_{\kappa lm}^\nu&=&(-1)^{l+\kappa+\nu}D_{\kappa lm}^\nu .
\end{eqnarray}
The superscript $\nu$ takes on values $\nu=0$ and $1$ while
the integer subscript $\kappa$ obeys $0 \le \kappa \le l$.
The quantum number $\kappa$ is the absolute value of an
angular momentum-like quantum number about the body-fixed axis
of rotation.
Even/odd $\kappa$ determines the parity eigenvalue while
the combination $l+\kappa+\nu$ determines the exchange eigenvalue. 
 This basis is especially useful since
each of the four operators above commutes with the atomic Hamiltonian, $\hat{H}_0$.
The spatial eigenfunction $\psi_{klms}[\bd{r}_1,\bd{r}_2]$ for
total spin $s$, total angular momentum $l$, 
$z$-component of angular momentum $m$, and parity $k=\pm 1$ satisfies
\begin{eqnarray}
\hat{L}^2\psi_{klms}[\bd{r}_1,\bd{r}_2] &=& l(l+1)\psi_{klms}[\bd{r}_1,\bd{r}_2]\\
\hat{L}_z\psi_{klms}[\bd{r}_1,\bd{r}_2] &=& m\psi_{klms}[\bd{r}_1,\bd{r}_2]\\
\hat{\Pi}\psi_{klms}[\bd{r}_1,\bd{r}_2] &=& k\psi_{klms}[\bd{r}_1,\bd{r}_2]\\
\label{PsiEigenStates}
\hat{{\cal E}}_{12}\psi_{klms}[\bd{r}_1,\bd{r}_2] &=& (-1)^s\psi_{klms}[\bd{r}_1,\bd{r}_2].
\end{eqnarray}
Equations \ref{DEigenStates}-\ref{PsiEigenStates} imply
\begin{equation}
\label{WaveFunction}
\psi_{klms}[\bd{r_1},\bd{r_2}]=\sum_{\nu=0}^1
\sideset{}{'} \sum_{\kappa=\nu}^l
g_{\kappa ls}^\nu[r_1,r_2,\theta_{12}]D_{\kappa lm}^\nu[\Theta,\Phi,\Psi],
\end{equation}
where the prime on the sum means that
$\kappa$ is restricted to even ($k=1$) or odd ($k=-1$) numbers if parity
is even or odd, 
respectively, and $g_{\kappa ls}^\nu$ is a real function of the internal coordinates. 
The convenience of the Bhatia and Temkin \cite{BhatiaTemkin1964}
coordinate choice is most evident in how one imposes total antisymmetry
of the wave function.
The spin singlet (triplet) must have a symmetric (antisymmetric) spatial wave function. 
The properties of the $D_{\kappa lm}^\nu$ functions reduce
this requirement to
\begin{equation}
\label{SpatialSymmetry}
\hat{{\cal E}}_{12}g_{\kappa ls}^\nu=(-1)^{\nu+\kappa+l+s}g_{\kappa ls}^\nu.
\end{equation}
The total antisymmetry of a wave function with given $k$,
$l$, $m$ and $s$ follows by imposing the above requirement under $r_1
\leftrightarrow r_2$ on each radial function
for each $\nu$ and $\kappa$. Note that
$(-1)^{\kappa + l + s}$ is fixed directly by the wave function's
$k$, $l$ and $s$. The same requirement applies to both singlet and
triplet states up to the difference in the value of $s$.

The full 
six-dimensional Schr\"{o}dinger equation for given $l$, $s$, even/odd parity,
and any $m$ yields $l$ or $l+1$ (depending on these quantum numbers) coupled three-dimensional equations for $g_{\kappa ls}^\gamma$. The indices for $g$ satisfy
$\gamma$ = $0$ or $1$ and $0 \le \kappa \le l$ with even or odd $\kappa$ for
even or odd parity. The equations are
\begin{equation}
0 = (\hat{H}_S-E)g_{\kappa ls}^\gamma+\sum_{\nu=0}^1\sum_{n=-1}^1\hat{H}_{\nu \kappa n}^\gamma g_{\kappa+2n,l,s}^\nu ,
\end{equation}
where $\hat{H}_S$ is the part of the Hamiltonian operator that survives for
S states. The summation enumerates couplings with
$\gamma \ne \nu$ and/or different $\kappa$ as well as terms that
are intrinsic to non-S-states.

Appendix \ref{BTAppendix} gives the explicit forms of
the operators $\hat{H}_S$ and $\hat{H}_{\nu \kappa
  n}^\gamma$.

\section{Review of the oscillator strength and dipole radiative transitions}
\label{DipoleReview}

The oscillator strength quantifies the coupling between
two eigenstates of $\hat{H}_0$ on account of
interactions with a perturbing electromagnetic field.
It is fundamental for interpreting spectra, including
the strength and width of atomic transitions and the
lifetimes of atomic states. Sites generating spectra of
interest are ubiquitous. They include earth-based
laboratories, photospheres of the Sun and distant
stars, and the near vacuum between the stars where
traces of interstellar matter radiate. The specific
applications of the oscillator strength are
correspondingly diverse.  For example, in laboratories
the technique of laser spectroscopy is used to measure
energy splittings and frequency-dependent
photoabsorption cross sections of highly excited
states. Knowledge of the transition probability
matrices is needed to interpret which states have been
directly and indirectly generated. The transitions
are driven by collisional and radiative processes, the
latter given in terms of oscillator strengths.  In an
astrophysical context, on the other hand, observations
of stellar emission require oscillator strengths for
inferring chemical abundances from absorption or
emission of radiation
\cite{Smith1973,BiemontGrevesse1977}. Oscillator
strengths have widespread utility.

The practical difficulty in calculating the oscillator
strength value is the accurate representation of the
initial and final wave functions.  Almost from the very
beginning of the development of quantum mechanics
helium, having but two electrons, has served as a
testing ground for new theoretical approaches. Appendix
\ref{OscStrengthTable}  presents a brief, schematic
description of the rich history of such improvements in
the service of oscillator strength calculations.

Following
Baym \cite{Baym1969} and Bethe and Salpeter \cite{BetheSalpeter1957},
the nonrelativistic Hamiltonian of a two-electron atom in the presence of an 
electromagnetic field (infinite-nuclear-mass approximation) is
\begin{equation}
\label{eq:hamiltonian}
\hat{H}_{EM}=\hat{H}_0+\hat{H}_{\textrm{int}},
\end{equation}
where $\hat{H}_0$ is the Hamiltonian for the isolated
atom (Eq. \ref{Hamiltonian0})
and $\hat{H}_{\textrm{int}}$ describes the interaction of the atom with radiation,
\begin{equation}
\hat{H}_{\textrm{int}} = \sum_{i} \left(
-\frac{\bd{p}_i \cdot \bd{A}_i 
+ \bd{A}_i \cdot \bd{p}_i}{2c} - \frac{A_i^2}{2c^2} + \varphi_i 
\right),
\label{eq:hint}
\end{equation}
where $\bd{A}_i$ and $\varphi_i$ are the vector and scalar potential, respectively, 
at the location of the $i$th electron (excluding the 
atomic Coulomb interactions included in $V$),
and $c$ is the speed of light.
If the photon number density is 
small then the second term, corresponding to two-photon processes, is much smaller than the 
first and if one adopts the transverse gauge then the third term is zero. 
With these assumptions the non-zero terms are the ones
linear in the vector potential.

Only electric dipole-mediated transitions and the associated $f$'s
are considered in this article.
The length, velocity and acceleration
forms for the oscillator strength \cite{Hibbert1975} are
\begin{eqnarray}
\label{OscStrengthEq}
f_{ij}^l&=&\frac{2}{3}(E_j-E_i)|\langle j|\bd{R}|i\rangle|^2\\
f_{ij}^v&=&\frac{2}{3}\frac{1}{E_j-E_i}|\langle j|\bd{P}|i\rangle|^2\\
f_{ij}^a&=&\frac{2}{3}\frac{1}{(E_j-E_i)^3}\left|\left\langle j\left|\bd{A}
\right|i\right\rangle\right|^2 .
\end{eqnarray}
Here $E_i$ and $E_j$ are the energies of the initial and final states.
The two-particle operators are
\begin{eqnarray}
\bd{R}& =& \bd{r_1}+\bd{r_2} \\
\bd{P}& =& \bd{p_1}+\bd{p_2} \\
\bd{A}& =& -\frac{Z\bd{r_1}}{r_1^3}-\frac{Z\bd{r_2}}{r_2^3} ,
\end{eqnarray}
i.e. the position, momentum and acceleration electron operators. 
Appendix \ref{BTOscStrength} presents explicit expressions for $f$ used
in the calculations.

If the wave functions, energies, and operators were exact, all three forms would
give identical results. However, in a numerical calculation the
agreement may be destroyed whenever the operator commutator rule
\begin{equation}
\bd{P}=i[\hat{H}_0,\bd{R}]
\end{equation}
is violated. Approximations to 
the operators ($\hat{H}_0$, $\bd{P}$, or $\bd{R}$) and to
the initial and final eigenstates are possible sources of error.
Good agreement between the 
three forms at a fixed resolution has sometimes been taken to be
an indication of an accurate answer. Such
agreement is ultimately necessary as resolution improves
but the closeness of the agreement is insufficient to infer the accuracy
at a fixed resolution \cite{SchiffEtAl1971,Hibbert1975}.
A more stringent approach involves two steps: first,
for each form check that
the matrix element
converges with resolution or basis size and, second, that
the converged answers for different forms agree.

The oscillator strengths $f_{0n}$ for transitions, $1^1$S $\to n^1$P
of helium obey a family of sum rules. For integer $k$ define
\begin{equation}
\label{SDefinition}
S(k) \equiv \sum_n|\Delta E_{0n}|^k f_{0n}.
\end{equation}
where the summation is over all P states, including the
continuum. Here, $\Delta E_{0n}$ is the energy difference with
respect to the ground state.
The rules \cite{DalgarnoLynn1957,Drake1996} include
\begin{eqnarray}
\label{SumRulesA}
S(-1)&=&\frac{2}{3}\langle(\bd{r}_1+\bd{r}_2)^2\rangle,\\
S(0)&=&2,\\
\label{SumRulek1}S(1)&=&-\frac{4}{3}\langle \hat{H}_0-\bd{p}_1\cdot\bd{p}_2\rangle,\\
\label{SumRulesB}
S(2)&=&\frac{2\pi Z}{3}\langle \delta(\bd{r}_1)+\delta(\bd{r}_2)\rangle,
\end{eqnarray}
where the expectation values on the right hand side refer to the
ground state.

In principle, these sum rules provide consistency checks 
on theoretically calculated oscillator strengths. However, the
explicit evaluation of $S(k)$ (Eq. \ref{SDefinition}) is
difficult. Multiple methods are needed to handle all the final
states, which include a finite number of low energy
highly correlated states, a countably infinite number
of highly excited states, and an uncountably infinite
number of continuum states. Ref. \cite{Berkowitz1997} inferred that
the two sides of Eqs. \ref{SumRulesA}-\ref{SumRulesB}
agree to about one percent based on a combination of
the most reliable theoretical and/or experimental values for $f_{0n}$.

This article exemplifies the capabilities of the
pseudospectral approach by evaluating the $1^1$S $\to
2^1$P oscillator strength, a physical regime in which
strong electron correlations are paramount, and a set
of expectation values for operator forms, some of which appear on
the right hand side of the sum rules.

\section{\label{VarNDom}Variables and Domains}
\label{VarsAndDomains}

This section details an important
element of the application of the pseudospectral method:
the choice of coordinates and computational
domains.

To achieve exponentially fast convergence with a pseudospectral
method, it is imperative that the solution be smooth. The presence of
a singular point may require a special coordinate choice in the
vicinity of the singularity or a different choice of effective basis.  
Handling multiple singularities
typically requires several individual subdomains, each accommodating
an individual singularity. It is useful to have a guide for choosing
appropriate coordinates.

The ordinary differential equation
\begin{equation}
\label{OneDimRegDifEq}
\left(
\fulldd{}{\gencoord}+\frac{p_a[\gencoord]}{\gencoord-a}\fulld{}{\gencoord}+\frac{q_a[\gencoord]}{(\gencoord-a)^2}\right)f=0
\end{equation}
with $p_a[\gencoord]$ and $q_a[\gencoord]$ analytic at $\gencoord=a$
has a regular singular point at $\gencoord=a$. The basic theory of ordinary differential equations
(ODE's) \cite{Coddington1961} states that
$f$ has at least one Frobenius-type solution about $\gencoord=a$ of the form
\begin{equation}
f[\gencoord]=(\gencoord-a)^{t_a}\sum_{n=0}^\infty c_n (\gencoord-a)^n,
\end{equation}
where the coefficients $c_n$ can be derived by directly plugging into Eq. \ref{OneDimRegDifEq} and
$t_a$ is the larger of the two solutions to the indicial equation
\begin{equation}
t_a(t_a-1)+p_a[a]t_a+q_a[a]=0.
\end{equation}
Exponential convergence of the pseudospectral method for a
differential equation of the form of Eq. \ref{OneDimRegDifEq} requires
$t_a$ be a non-negative integer.  This must hold at each singularity
$a$ in the domain (as well as all other points).\footnote{The full class of one dimensional problems for which
pseudospectral methods converge exponentially fast
is larger than this description. 
The method needs the solution to be smooth which is a 
weaker statement than that it
be analytic. This 
distinction is not material for the 
singular points discussed here.}

A simple example is the Schr\"{o}dinger equation for a
hydrogenic atom expressed in spherical coordinates 
$\{\gencoord_1,\gencoord_2,\gencoord_3\}=\{r,\theta,\phi\}$.
The radial part of the full wave function $R_{nl}[r]$ satisfies
\begin{equation}
\left(\fulldd{}{r}+\frac{2}{r}\fulld{}{r}-\frac{l(l+1)-2Zr-2Er^2}{r^2}\right)R_{nl}=0.
\end{equation}
A comparison with Eq. \ref{OneDimRegDifEq} yields $p_0[0]=2$ and
$q_0[0]=-l(l+1)$, which gives $t_0=l$, the well known result for
hydrogenic wave functions.
The reduction of the partial differential equation (PDE) into an ODE having non-negative integer $t_0$
tells us that spherical coordinates are a good choice for solving hydrogenic wave functions
using pseudospectral methods. A bad choice would be Cartesian coordinates
$\{\gencoord_1,\gencoord_2,\gencoord_3\}=\{x,y,z\}$.
The ground state has the form
\begin{equation}
\psi \propto e^{-Z\sqrt{x^2+y^2+z^2}}.
\end{equation}
This solution has a discontinuity in its first derivatives at $x=y=z=0$:
\begin{equation}
\lim_{x,y,z\to 0^+}\frac{\partial\psi}{\partial x,y,z}\ne\lim_{x,y,z\to 0^-}\frac{\partial\psi}{\partial x,y,z}.
\end{equation}
Other solutions have a discontinuity of first or higher derivatives at the same point.
The pseudospectral method would not
handle these well and convergence would be limited to being algebraic.

An arbitrary second order PDE may have
singularities that occur on complicated hypersurfaces of different
dimensionality. Deriving the analytic properties of a solution near such
a surface is a daunting task. The general idea is to
seek a coordinate system such that the limiting
form of the PDE near the singularity looks like an ODE of the sort
that pseudospectral methods are known to handle well.

For example, in a three-dimensional space, assume the singularity lies
on a two-dimensional surface. First, seek a coordinate system such
that the surface occurs at $\gencoord_1=a$.\footnote{A zero- or one-dimensional singularity
can be made to look two-dimensional by a coordinate transformation. For example,
in the previous example, which has used spherical coordinates, the Coulomb 
singularity appears at $r=0$. This point is approached on a two-dimensional sphere of constant radius by
taking the limit as a single coordinate, the radius, approaches zero.}
Second, focusing on
$\gencoord_1$, seek coordinates so that is possible to rewrite the PDE
in the form
\begin{equation}
\label{NDimRegDifEq}
\left(\pdd{\gencoord_1}+\frac{\hat{P}_{a}[\bd{\gencoord}]}{\gencoord_1-a}\pd{\gencoord_1}+\frac{\hat{Q}_{a}[\bd{\gencoord}]}{(\gencoord_1-a)^2}\right)f=0
\end{equation}
where $\hat{P}_{a}$ and $\hat{Q}_{a}$ are linear second order
differential operators that do not include derivatives with respect to
$\gencoord_1$.  Finally, seek coordinates such that
$\hat{P}_{a}$ and $\hat{Q}_{a}$ are analytic
with respect to $\gencoord_1$ at $a$. 

Unfortunately, even if one succeeds in finding such a coordinate
system, the
theorem of ODEs does not generalize to PDEs, i.e. there is no guarantee
that $f$ is analytic near $a$. A celebrated example is exactly the
problem of concern here, i.e. the Schr\"{o}dinger equation for two-electron
atoms. Three coordinates are needed to describe the S state.
In hyperspherical coordinates ($\{X_1, \dots\}=\{\rho, \dots\}$ where
$\rho = \sqrt{r_1^2 + r_2^2}$), Schr\"{o}dinger's equation matches the form of
Eq. \ref{NDimRegDifEq} for $\gencoord_1=\rho$ and $a = 0$. This is the triple
coalescence point, a point singularity in the three-dimensional
subspace spanned by the coordinates $r_1$, $r_2$, and $r_{12}$. 
The electron-nucleus and electron-electron singularities (two-body
coalescence points)
are one-dimensional lines in this subspace that meet at $\rho=0$. 
Bartlett \cite{Bartlett1937}
proved that no wave function of the form
\begin{equation}
\psi = \sum_{n=0}^\infty A_n \rho^n,
\end{equation}
where $A_n$ is an analytic function of the remaining variables
will satisfy the PDE. Fock's form for the solution \cite{Fock1954,Fock1958} is
\begin{equation}
\psi = \sum_{n=0}^\infty\sum_{m=0}^{\lfloor n/2 \rfloor}B_{nm}\rho^n(\log\rho)^m,
\end{equation}
where $B_{nm}$ is an analytic function of the remaining variables. The
presence of the $\log \rho$ terms in the wave function is an important 
qualitative distinction between a solution having
two- and three-body coalescence points.

Some properties of the solution near $\rho=0$ have been reviewed in
our previous article \cite{GrabowskiChernoff2010}. For example, Myers {\it et al.}
\cite{MyersEtAl1991} showed that the logarithmic terms allow the local energy $(\hat{H}\psi)/\psi$
near $\rho=0$ to be continuous.  Despite this property, they have only a
slight effect on the convergence of variational energies
\cite{Schwartz2006}. By many measures of error the triple coalescence
point does not affect pseudospectral calculations until very high
resolutions \cite{GrabowskiChernoff2010}.  


As a point of principle, however,
no simple coordinate choice can hide the problems that occur at the triple
coalescence point, and no special method for handling this singularity is given here. 
Elsewhere ($\rho \ne 0$) our rule of thumb is the
following: coordinates are selected so that
the singularity may be described by 
$X_i=a$ with $\hat{P}_{a}$ and $\hat{Q}_{a}$  satisfying
\begin{eqnarray}
\label{PCondition}
\hat{P}_{a}&=&\sum_{n=0}^\infty (\gencoord_i-a)^n \hat{p}_{an}\\
\label{QCondition}
\hat{Q}_{a}&=&\sum_{n=0}^\infty (\gencoord_i-a)^n \hat{q}_{an},
\end{eqnarray}
in a neighborhood about $\gencoord_i = a$. Here, $\hat{p}_{an}$ and $\hat{q}_{an}$ 
are linear differential operators not containing $\gencoord_i$ or 
its derivatives.

The singularities of the Hamiltonian, given in detail in
Appendix \ref{BTAppendix}, are of two types.  The physical
singularities at $r_1$, $r_2$, and $r_{12}=0$ were explored in
Ref. \cite{GrabowskiChernoff2010}. One of the essential virtues of
hyperspherical coordinates is that
$\rho \ne 0$ implies these coalescences have separate neighborhoods.
Therefore, the prescription is to seek separate coordinates satisfying
eqs. \ref{PCondition} and \ref{QCondition}
in the vicinity of each singularity.

There are also coordinate singularities at $\theta_{12}=0$ and $\pi$
which correspond to collinear arrangements of the two electrons and
nucleus. These singularities were completely
absent in our previous treatment of S states 
\cite{GrabowskiChernoff2010} where
$C=-\cos\theta_{12}$ and $B=-\cos\beta_{12}$ ($\beta_{12}$ is defined below) 
were the third coordinates in different subdomains. Now, to accommodate
the singularities' presence in the Hamiltonian for general angular momentum
make the slight change to use $\theta_{12}$ and $\beta_{12}$ instead.

Starting with the internal coordinates $r_1$, $r_2$ and $\theta_{12}$
one defines $\rho$, $\phi$, $\zeta$, and $x$ by
\begin{eqnarray}
r_1 & = & \rho \cos{\phi}\\
r_2 & = & \rho \sin{\phi}\\
r_{12}&=& \rho \sqrt{2}\sin\zeta\\
\sqrt{2}\sin{\zeta} & = & \sqrt{1-\cos\theta_{12}\sin{2\phi}}\\
\cos\beta_{12} & = &-\frac{\cos{2\phi}}{\sqrt{1-\cos^2\theta_{12}\sin^2{2\phi}}}\\
x & = &\frac{1-\rho}{1+\rho}.
\end{eqnarray}
The full ranges of these variables are
\begin{equation}
\label{varRanges}
\begin{array}{c}
0\le r_1,r_2,\rho < \infty\\ 
|r_1-r_2|\le r_{12} \le r_1+r_2\\
0\le \theta_{12},\beta_{12} \le \pi\\
0\le \phi,\zeta\le \pi/2 \\
-1\le x\le 1.
\end{array}
\end{equation}
The purpose of coordinate $x$ is to map the semi-infinite range of
$\rho$ to a finite interval.

\begin{figure}
\includegraphics[width=\linewidth]{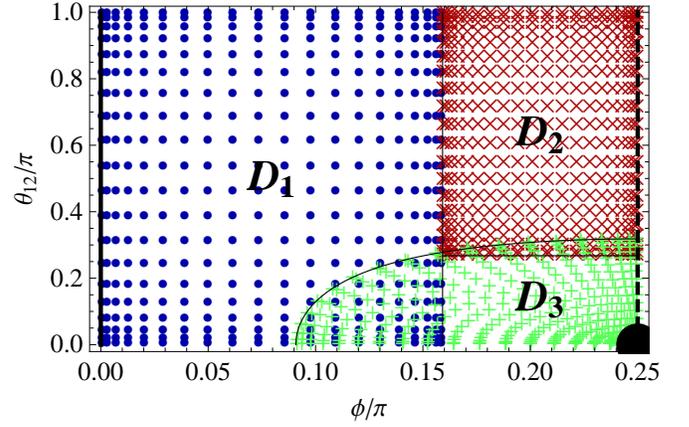}%
\caption{\label{GridPoints} 
(Color online).
This is the arrangement of grid points of the three domains at a
constant value of $\rho$ in $\phi$ and $\theta_{12}$ coordinates for $n=20$. Note that the
point density becomes larger at the boundary of each subdomain and
that no grid points sit on the Coulomb singularities. The blue circles,
red crosses, and green pluses belong to domains $D_1$, $D_2$, and $D_3$,
respectively. $D_1$ and $D_2$ are rectangular domains, while $D_3$ has the 
curved boundary in $\phi$, $\theta_{12}$ coordinates but is rectangular in $\zeta$, $\beta_{12}$ coordinates. 
The electron-proton singularity occurs on the left side
(solid line at $\phi=0$). The entire line corresponds to one physical
point. The electron-electron singularity occurs at the lower right hand
corner (solid disk at $\phi=\pi/4, \theta_{12}=0$). A line of symmetry falls on
the right side (dashed line at $\phi=\pi/4$ where $r_1=r_2$).}
\end{figure}

Eqs. \ref{PCondition} and \ref{QCondition} are satisfied by
selecting $\{X_1,X_2,X_3\}=\{x,\phi,\theta_{12}\}$ or $\{x,\zeta,\beta_{12}\}$
in three separate domains
\begin{equation}
\begin{array}{lccc}
D_1:&-1\le x\le 1, & 0\le\phi\le\frac{1}{2}, & -1\le \cos\theta_{12}\le 1\\
D_2:&-1\le x\le 1, & \frac{1}{2}\le\phi\le\frac{\pi}{4}, & -1\le\cos\theta_{12}\le\frac{2}{3}\\
D_3:&-1\le x\le 1, & 0\le\zeta\le\frac{1}{2}, & -1\le \cos\beta_{12}\le 0,
\end{array}
\end{equation}
spanning only half the space defined by the inequalities (\ref{varRanges}) due to
the symmetry in the Hamiltonian about $r_1=r_2$.
Fig. \ref{GridPoints} illustrates the layout of the 
three domains at fixed $\rho$. The coordinate systems in domains
$D_1$ and $D_3$ were developed to handle the electron-proton and 
electron-electron singularities, respectively. The choice of coordinates
in domain $D_2$ was more arbitrary, and for simplicity was chosen to be 
the same as in domain $D_1$. This particular choice allows for no overlap 
between domains $D_1$ and $D_2$ and makes the symmetry condition (Eq. \ref{SpatialSymmetry})
at $r_1=r_2$, $\phi=\pi/4$, or $\beta_{12}=\pi/2$ easy to apply.
The remaining electron-nucleus
singularity, $r_1=0$, is implicitly accommodated by the spatial
symmetry of the wave function.
The three domains must jointly describe the
full rectangle but the specific choice for edges at $\phi=\zeta=1/2$ is
arbitrary.

\section{Boundary conditions}
\label{BoundaryConditions}
\subsection{Internal boundary conditions}

It is necessary to ensure continuity of
the wave function and its normal derivative at internal boundaries.
There are two ways in which the subdomains can touch:
they can overlap or they can barely touch.
For clarity, consider a one-dimensional problem with two domains.
Let the first domain be domain $1$ and the second be 
domain $2$ with extrema 
$\gencoord_{1,\textrm{min}}<\gencoord_{2,\textrm{min}}\le \gencoord_{1,\textrm{max}}<\gencoord_{2,\textrm{max}}$,
where the $1$ and $2$ refer to domain number.
The first case corresponds to $\gencoord_{2,\textrm{min}}<\gencoord_{1,\textrm{max}}$ 
and the second to $\gencoord_{2,\textrm{min}}=\gencoord_{1,\textrm{max}}\equiv \gencoord_*$.
For both cases, exactly two conditions are needed to make the wave function
and its derivative continuous. The simplest choice for the first case is
\begin{eqnarray}
\psi_1[\gencoord_{1,\textrm{max}}]&=&\psi_2[\gencoord_{1,\textrm{max}}]\\
\psi_1[\gencoord_{2,\textrm{min}}]&=&\psi_2[\gencoord_{2,\textrm{min}}],
\end{eqnarray}
and for the second case is
\begin{eqnarray}
\psi_1[\gencoord_{*}]&=&\psi_2[\gencoord_{*}]\\
\frac{d}{d\gencoord}\psi_1[\gencoord_{*}]&=&\frac{d}{d\gencoord}\psi_2[\gencoord_{*}].
\end{eqnarray}

For multi-dimensional grids, the situation is analogous.
The conditions are applied 
on surfaces of overlap. In this case
the derivatives are surface normal derivatives
or any derivative not parallel to the boundary surface.
On a discrete grid, a finite number of conditions
are given which, in the limit of an infinitely fine mesh,
would cover the entire surface. Additional discussion
and illustrations of the technique are in
Ref. \cite{GrabowskiChernoff2010}.

\subsection{The symmetry condition}

The Hamiltonian (see appendix \ref{BTAppendix}) is symmetric with
respect to particle exchange ($r_1\leftrightarrow r_2$). Therefore, there are two types of
eigenstates: those with symmetric spatial wave functions (singlets)
and those with antisymmetric spatial wave functions (triplets). The
radial wave functions $g_{\kappa ls}^\nu$ satisfying the appropriate
symmetry must obey Eq. \ref{SpatialSymmetry}. More explicitly
\begin{equation}
0=\left\{\begin{array}{cc}
\left.\partd{g_{\kappa l s}^\nu}{\phi}\right|_{\phi=\pi/4}=\left.\partd{g_{\kappa l s}^\nu}{\beta_{12}}\right|_{\beta_{12}=\pi/2}
& \textrm{ if } \xi \textrm{ is even}\\
\left.g_{\kappa l s}^\nu\right|_{\phi=\pi/4}=\left.g_{\kappa l s}^\nu\right|_{\beta_{12}=\pi/2}
& \textrm{ if } \xi \textrm{ is odd}
\end{array}\right.,
\end{equation}
where $\xi=\nu+\kappa+l+s$.

\section{Energy and oscillator strength results}
\label{EandOscResults}
This article generalizes the pseudospectral methods previously
developed for S states to the general angular momentum case,
calculates oscillator strengths for transitions, and tests how
different measures of wave function errors vary with resolution.

\begin{figure}
\includegraphics[width=\linewidth]{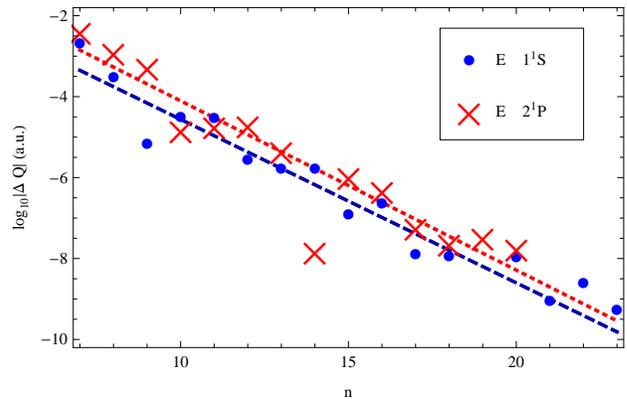}
\caption{\label{Energy} 
(Color online).
The logarithm base 10 of the energy error ($\Delta Q$) of both the lowest energy S state and P 
state of helium. The dark blue circles are for the $1^1$S state
and the light red crosses for the $2^1$P state with dashed blue and dotted red fits, respectively (see Tab. 
\ref{ConvTabEOS}).}
\end{figure}

The most widely quoted number to ascertain convergence is the energy which
gives a global measure of accuracy. Figure \ref{Energy} shows the energy errors for 
the 1$^1$S and 2$^1$P states of helium. Here and throughout the results sections the high precision values 
of  Drake \cite{Drake1996} are taken to be exact. The energy error for
both states decreases exponentially with resolution. Convergence for the S
state is similar to that reported in Ref. \cite{GrabowskiChernoff2010}
with slight differences related to
a different choice of coordinates. The current calculation extends
to basis size $n=23$ for S states and $n=20$ for P states instead of $n=14$ for only S states in 
Ref. \cite{GrabowskiChernoff2010}.

A common feature of the energy convergence and all other convergence plots in this article is 
non-monotonic convergence. This method is not variational, so there is no reason to expect monotonic
convergence. Calculated quantities can fall above or below their actual value, with error
quasi-randomly determined by the exact grid point locations. The 
jumps decrease in magnitude as the resolution is increased. 

\begin{figure}
\includegraphics[width=\linewidth]{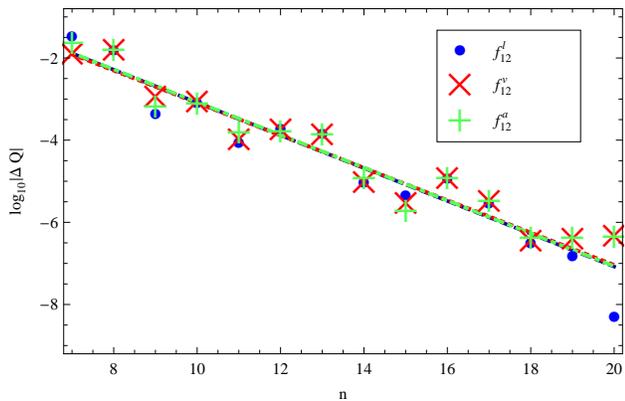}
\caption{\label{OscStrength} 
(Color online).
The logarithm base 10 of the error ($\Delta Q$) in the oscillator strength of the 
$1^1$S $\to 2^1$P transition of helium. The dark blue circles are for the length form,
the light red crosses are for the velocity form,  and the green pluses for 
the acceleration form with dashed blue, dotted red, and dot-dashed green fits, 
respectively (see Tab. \ref{ConvTabEOS}).}
\end{figure}

As described in Sec. \ref{DipoleReview}, there are three commonly used
forms for the oscillator strength. The length, velocity, and
acceleration forms depend most strongly on the value of the wave
function at positions in configuration space corresponding to large,
medium, and small separations. Sometimes the relative errors are used to
infer where the wave function is more or less accurate.
It has been observed that
for most variational calculations, the acceleration form tends to be
much less accurate than the other two forms, suggesting errors in the
wave function at small separation that have little effect on the
variational energy.  The length and velocity forms give
results of roughly comparable accuracy.

The oscillator strength of the $1^1$S $\to 2^1$P transition was
calculated using all three forms and Fig. \ref{OscStrength} displays
the errors.
Here, all three forms give roughly the same results. 
At most resolutions the points lie nearly 
on top of one another and their fits are indistinguishable,
indicating the wave function errors for small, medium, and 
large separations have roughly equal contributions to the numerically
calculated oscillator strength.
This may be due to the pseudospectral method's equal treatment of all 
parts of configuration space.

It should be noted that the value used as the exact value
\cite{Drake1996} is given to seven decimal places. 
Consequently, the errors inferred for the highest resolution 
calculations in Fig. \ref{OscStrength} are
not too precise. There is little practical need for additional
digits since a host of other effects including finite nuclear
mass, relativistic, and quadrupole corrections would confound
any hypothetical, experimental measurement
of the oscillator strength to such high precision
even if a perfect measurement could be made.
Actual experiments struggle to obtain two percent precision \cite{ZitnikEtAl2003},
an error larger than these effects.

\begin{figure}
\includegraphics[width=\linewidth]{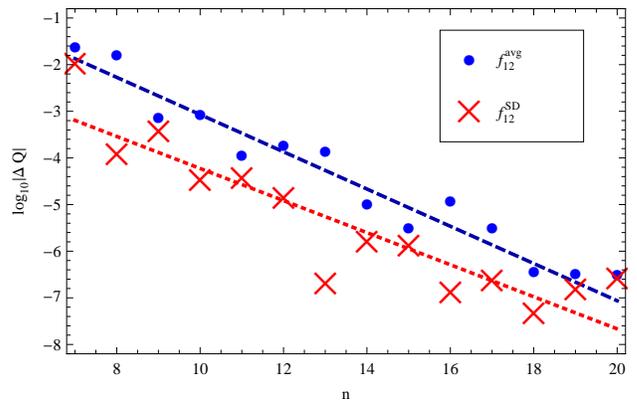}
\caption{\label{OSStat} 
(Color online).
The logarithm base 10 of the error ($\Delta Q$) in the average of the length, velocity, 
and acceleration forms of the oscillator strength (dark blue circles)  and their standard deviation 
(light red crosses) for the 
$1^1$S $\to 2^1$P transition of helium, with dashed blue and dotted red fits, 
respectively (see Tab. \ref{ConvTabEOS}).}
\end{figure}

As pointed out by Schiff {\it et al.} \cite{SchiffEtAl1971} and
reviewed by Hibbert \cite{Hibbert1975}, the assumption that using the
differences between the oscillator strength values from the
different forms as a measure of the accuracy is not
valid. Agreement is necessary but not sufficient.
They suggest comparing
calculated and extrapolated values. This latter procedure is
not straightforward for a pseudospectral method with
non-monotonic convergence. We present a similar suitable check.
Fig. \ref{OSStat} shows the average and
standard deviation of the error for the three forms as a function of
resolution.  The standard deviation is about an order of magnitude (with a 
large scatter about that factor of ten)
less than the average error at low and moderate resolutions
but the trend lines suggest that the standard deviation may be
approaching the average at the higher resolutions.
A possible explanation is that the calculation at the highest resolutions
is starting to become sensitive to the wave function truncation
(see appendix \ref{MatEigSol}). This destroys the expected
equality between the forms and 
each form converges to its own incorrect asymptotic value. The
individual errors and the standard deviation become comparable.
So at $n=20$, we assume the standard deviation and total error are equal
and get a value for the oscillator strength of 
$0.27616499(27)$ which compares
favorably to Drake's $0.2761647$ \cite{Drake1996}.

\begin{table}
\caption{\label{ConvTabEOS} The fit parameters to all the convergence plots of quantities $Q$ in this section.}
\begin{ruledtabular}
\begin{tabular}{|c|c|c|c|}
$Q$ & Figure & $A$ & $\beta$\\
\hline
$E(1^1$S$)$ & \ref{Energy} & $2.5\times 10^{-9}$ & 0.40\\
$E(2^1$P$)$ & \ref{Energy} & $5.2\times 10^{-9}$ & 0.42\\
$f_{12}^l$ & \ref{OscStrength} & $8.4\times 10^{-8}$ & 0.40\\
$f_{12}^v$ & \ref{OscStrength} & $9.2\times 10^{-8}$ & 0.39\\
$f_{12}^a$ & \ref{OscStrength} & $8.6\times 10^{-8}$ & 0.40\\
$f_{12}^{avg}$ & \ref{OSStat} & $8.7\times 10^{-8}$ & 0.40\\
$f_{12}^{SD}$ & \ref{OSStat} & $2.2\times 10^{-8}$ & 0.34
\end{tabular}
\end{ruledtabular}
\end{table}

All convergence data were fit to functions of the form $\Delta Q = A\times 10^{-\beta(n-20)}$ using the same procedure as 
in Ref. \cite{GrabowskiChernoff2010}. Because of uncertainty in the errors for the largest
resolutions ($n=19$ and $n=20$) these points were not used in the fits of $f_{12}^l$, $f_{12}^v$,
$f_{12}^a$, and $f_{12}^{avg}$.
The $\beta$ parameter, which corresponds to the slope of the fits in the convergence graphs is roughly the
same for all fits, with the exception of the standard deviation of the oscillator strength forms. This 
behavior is consistent with our discussion of errors in the previous paragraph.

\section{Corrections to the Hamiltonian}
\label{CorrectionsHamiltonian}
Two small parameters appear in the full physical Hamiltonian: the ratio of the 
reduced mass of the electron-nucleus pair to the nuclear mass, $\mu/M=1.37074563559(58)\times 10^{-4}$
\cite{MohrEtAl2008a,MohrEtAl2008b} (for ${}^4$He)
and the fine structure constant $\alpha=7.2973525376(50)\times 10^{-3}$ \cite{MohrEtAl2008a,MohrEtAl2008b}.
Here, the lowest order corrections in $\mu/M$ and $\alpha$ are considered.
For very high-precision work, one needs the perturbative corrections 
in powers of each small quantity.

\subsection{Finite nuclear mass correction}
\label{FiniteNuclearMass}

The nonrelativistic ($\alpha^0$) Hamiltonian for two-electron atoms is
\begin{equation}
\hat{H}_{\textrm{nr}}=\hat{H}_0+\hat{H}_{\textrm{cm}}+\hat{H}_{\textrm{mp}},
\end{equation}
where $\hat{H}_0$ is the fixed-nucleus approximation to the Hamiltonian with the electron
mass set to $\mu$, $\hat{H}_{\textrm{cm}}$ is the
kinetic energy of the center of mass, and $\hat{H}_{\textrm{mp}}$ is the mass polarization term:
\begin{eqnarray}
\label{MassHamiltonian}
\hat{H}_0&=&\frac{1}{2}(p_1^2+p_2^2)+\hat{V}\\
\hat{H}_{\textrm{cm}}&=&\frac{1}{2(M+2m_e)}{p}_{\textrm{cm}}^2\\
\label{HMassPolarization}\hat{H}_{\textrm{mp}}&=&\frac{1}{M}\bd{p}_1\cdot\bd{p}_2,
\end{eqnarray}
where $\hat{V}$ is the potential energy operator, $m_e$ is the electron mass, $\bd{p}_{\textrm{cm}}$ is the 
momentum operator of the center of mass,  
and reduced mass atomic units ($\mu=1$) are being used.
The second term is removed in center-of-mass coordinates and the last term provides the dominant
nontrivial correction for finite nuclear mass (the trivial one being the scaling of the energy by 
$m_e/\mu$).

\subsection{Relativistic corrections}
\label{RelativisticCorrections}

The Schr\"{o}dinger equation is a nonrelativistic approximation to the true equation of motion. The lowest order relativistic corrections
enter at order $(\alpha^2)$, as summarized in Ref. \cite{Fischer1996} and repeated here. Note, all references in this article to orders in 
$\alpha$ are in Rydbergs.
The Breit-Pauli Hamiltonian encapsulates the correction
\begin{equation}
\hat{H}_{\textrm{BP}}=\hat{H}_{\textrm{nr}}+\hat{H}_{\textrm{rel}},
\end{equation}
where $\hat{H}_{\textrm{nr}}$ is the usual nonrelativistic Hamiltonian used in Schr\"{o}dinger's equation 
and $\hat{H}_{\textrm{rel}}$ is the lowest order relativistic correction. The latter can be further divided into
non-fine-structure (NFS) and fine-structure (FS) contributions:
\begin{eqnarray}
\hat{H}_{\textrm{NFS}}&=&\hat{H}_{\textrm{mass}}+\hat{H}_{\textrm{D}}+\hat{H}_{\textrm{SSC}}+\hat{H}_{\textrm{OO}}\\
\hat{H}_{\textrm{FS}}&=&\hat{H}_{\textrm{SO}}+\hat{H}_{\textrm{SOO}}+\hat{H}_{\textrm{SS}}.
\end{eqnarray}
The separate contributions to the Hamiltonian are
the mass-velocity (mass), two-body Darwin (D), spin-spin contact (SSC), orbit-orbit (OO), 
spin-orbit (SO), spin-other-orbit (SOO), and the spin-spin (SS) terms.
These are explicitly given by
\begin{eqnarray}
\label{HMassVelocity}\hat{H}_{\textrm{mass}}&=&-\frac{\alpha^2}{8}\sum_ip_i^4\\
\label{HDarwin}\hat{H}_{\textrm{D}}&=&-\frac{\alpha^2Z}{8}\sum_i\nabla_i^2r_i^{-1}+\frac{\alpha^2}{4}\sum_{i<j}\nabla_i^2r_{ij}^{-1}\\
\label{HSpinSpinContact}\hat{H}_{\textrm{SSC}}&=&-\frac{8\pi\alpha^2}{3}(\bd{s}_1\cdot\bd{s}_2)\delta(\bd{r}_{12})\\
\label{HOrbitOrbit}\hat{H}_{\textrm{OO}}&=&-\frac{\alpha^2}{2}\left(\frac{\bd{p}_1\cdot\bd{p}_2}{r_{12}}
+\frac{\bd{r}_{12}(\bd{r}_{12}\cdot \bd{p}_1)\cdot\bd{p}_2}{r_{12}^3}\right)\\
\hat{H}_{\textrm{SO}}&=&\frac{\alpha^2Z}{2}\sum_i\frac{\bd{\hat{l}}_i\cdot\bd{\hat{s}}_i}{r_i^3}\\
\hat{H}_{\textrm{SOO}}&=&-\frac{\alpha^2}{2}\sum_{i\ne j}\left(\frac{\bd{r}_{ij}}{r_{ij}^3}\times\bd{p}_i\right)
\cdot(\bd{s}_i+2\bd{s}_j)\\
\hat{H}_{\textrm{SS}}&=&\frac{\alpha^2}{r_{12}^3}\left(\bd{s}_1\cdot \bd{s}_2
-\frac{3}{r_{12}^2}(\bd{s}_1\cdot \bd{r}_{12})(\bd{s}_2\cdot \bd{r}_{12})\right),
\end{eqnarray}
where $i$ and $j$ can be 1 or 2, $\bd{p}_{i}$ and $\bd{r}_i$ are the momentum and position of the 
$i$th electron with respect to the nucleus, respectively,
$\bd{r}_{12}$ is the vector pointing from the first electron
to the second, and $\bd{\hat{s}}_i$ and $\bd{\hat{l}}_i$ are the one-electron spin and angular momentum operators 
of the $i$th electron, respectively. The last three Hamiltonian
terms are zero for ${}^1$S states due to 
symmetry considerations.

There are many higher order terms (see Refs. \cite{Drake1999,DrakeMorton2007,DrakeYan2008,PachuckiYerokhin2011}) 
but these are not considered here.

\section{Mass polarization and relativistic correction calculations}
\label{MassPolRelCorrCalcs}
The mass polarization and low order relativistic corrections to the nonrelativistic Hamiltonian
have been known for some time \cite{BetheSalpeter1957}. The main challenge in calculating these
terms is finding adequate unperturbed wave functions. Early calculations 
\cite{Kinoshita1957,KabirSalpeter1957,Sucher1958,ArakiEtAl1959} were critical
for comparing experimental and theoretical energies, confirming that Schr\"{o}dinger's equation
is correct in the nonrelativistic limit for helium.

The development of computers enabled Pekeris and coworkers 
\cite{Pekeris1958,Pekeris1959,SchiffEtAl1965} and others
\cite{Schwartz1961,Schwartz1964,Hambro1972,LewisSerafino1978,DavisChung1982}
to reach theoretical uncertainties in the energy of about $10^{-2}$ cm$^{-1}$. Such
precision and the resulting precision in the wave function 
allowed Lewis and Serafino \cite{LewisSerafino1978} to calculate the 
fine structure constant from experimental measurements of the $2^3$P splitting. 
They obtained $\alpha^{-1}=137.03608(13)$ with an estimated uncertainty
only surpassed at the time by
the measurements of the electron anomalous magnetic moment $(g-2)$ 
(by a factor of two) and
the ac Josephson experiments (by a factor of four).

Drake and collaborators \cite{Drake1987,DrakeYan1992,YanDrake1995,DrakeGoldman1999,Drake1999,Drake2002,Drake2004} 
and Pachucki and collaborators \cite{Pachucki1998,PachuckiSapirstein2000,PachuckiSapirstein2002,PachuckiSapirstein2003,Pachucki2006,Pachucki2006b,Pachucki2006c,PachuckiYerokhin2009,PachuckiYerokhin2010,PachuckiYerokhin2011,PachuckiYerokhin2011b} have pushed relativistic corrections for regular helium up to order $\alpha^5$ and 
beyond using a Hylleraas \cite{Hylleraas1929} type basis. Drake \cite{Drake2002} matched theoretical and 
observed energy differences in the $J=0,1$ splitting of the $2^3$P state 
and determined
$\alpha^{-1}=137.0359893(23)$. Drake cited a difference with the $g-2$ result 
$137.0359996(8)$
but agreement with the ac Josephson result $137.0359872(43)$ \cite{Drake2002}. 
However, a similar calculation of his using
the observed $J=1,2$ splitting gives an unreasonable value \cite{Drake2002}. 
Pachucki and collaborators have resolved the issue by finding errors in $\alpha^5$ terms and by 
increasing the error estimate due to $\alpha^6$ terms. Their most recent determination is
$\alpha^{-1}=137.03599955(64)(4)(368)$, where the first error is experimental, the second numerical,
and the third is their estimated error from higher order terms \cite{PachuckiYerokhin2011}. This value
agrees with the latest $g-2$ results but is not as precise \cite{PachuckiYerokhin2011}.

An alternative approach is to use an even simpler basis, with surprisingly accurate results. 
Korobov and collaborators have used an exponential basis (see Refs. \cite{ThakkarSmith1977,FrolovSmith1995}) 
to calculate very precise helium \cite{KorobovKorobov1999, Korobov2000,KorobovYelkhovsky2001,Korobov2002,Korobov2002b,Korobov2004}
(up to order $\alpha^4$) and anti-protonic helium \cite{KorobovEtAl1999,KorobovBakalov2001,Korobov2003,Korobov2006,KorobovTsogbayar2007,Korobov2008,KorobovZhong2009,KorobovZhong2009b}
(up to order $\alpha^5$) electronic energies. The latter calculations have been used for the
CODATA06 \cite{MohrEtAl2008a,MohrEtAl2008b} recommended value of the electron-to-(anti)proton mass ratio.

\section{Expectation values}
\label{ExpectationValues}
\begin{figure}[h!]
\includegraphics[width=\linewidth]{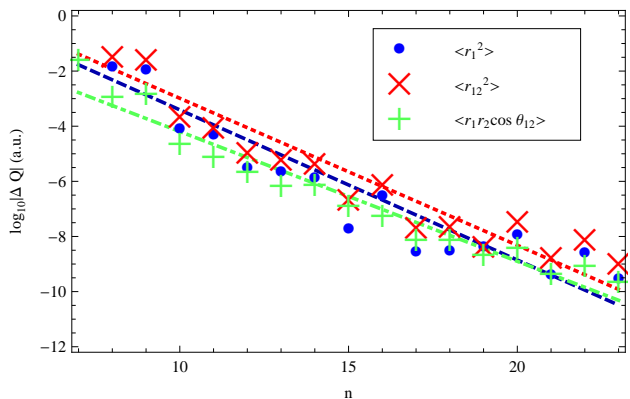}
\caption{\label{rSq} 
(Color online).
The logarithm base 10 of the error ($\Delta Q$) in the expectation values of operators that 
scale as $\rho^2$ for helium. The dark blue circles are for $\langle r_1^2\rangle$,
the light red crosses are for $\langle r_{12}^2 \rangle$, and the green pluses are for 
$\langle r_1 r_2\cos\theta_{12}\rangle$ with dashed blue, dotted red, and dot-dashed green fits, 
respectively (see Tab. \ref{ConvTab}).}
\end{figure}

The aim of this section is to test the pseudospectral method's ability
to represent the wave function in different parts of configuration
space and to compare the convergence rates of the errors with that of
the energies and oscillator strengths. For a representative set of
calculations consider the expectation values of the operators
needed for leading order relativistic
(Sec. \ref{RelativisticCorrections}) and finite nuclear mass
(Sec. \ref{FiniteNuclearMass}) corrections, for the oscillator
strength sum rules (Eqs. \ref{SumRulesA}-\ref{SumRulesB}), 
interparticle distances, $\langle \hat{V} \rangle$, and
$\langle \hat{V}^2 \rangle$.  These expectation values test different parts of the wave function
as well as different types of operators. They are organized by the weighting of
the wave function and used to draw inferences about local errors.


\begin{figure}[h!]
\includegraphics[width=\linewidth]{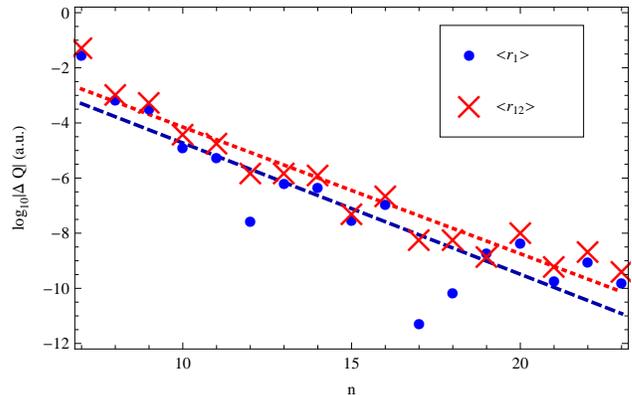}
\caption{\label{r} 
(Color online).
The logarithm base 10 of the error ($\Delta Q$) in the expectation values of operators that 
scale as $\rho$ for helium. The dark blue circles are for $\langle r_1\rangle$ and
the light red crosses are for $\langle r_{12}\rangle$ 
with dashed blue and dotted red fits, 
respectively (see Tab. \ref{ConvTab}).}
\end{figure}

Figure \ref{rSq} displays results for expectation values
related to sum rule $S(-1)$ (Eq. \ref{SumRulesA}), i.e. quantities scaling like $\rho^2$. 
These calculations are somewhat more sensitive to the wave function
at large separation than, say, the normalization integral. In addition, they
focus on parts of coordinate space which have low resolution
compared to the coverage near the singularities.
High accuracy is found for all three cases.

Figure \ref{r} displays results for expectation values
of operators scaling like $\rho$ similar to the length form of the oscillator strength. Higher accuracy is obtained here than for the oscillator
strength at equivalent resolutions. This can be explained by the smaller length scale set by the higher energy of the 
P state, which enters only into the oscillator strength calculations. So a greater resolution is 
needed for the same accuracy.


\begin{figure}[h!]
\includegraphics[width=\linewidth]{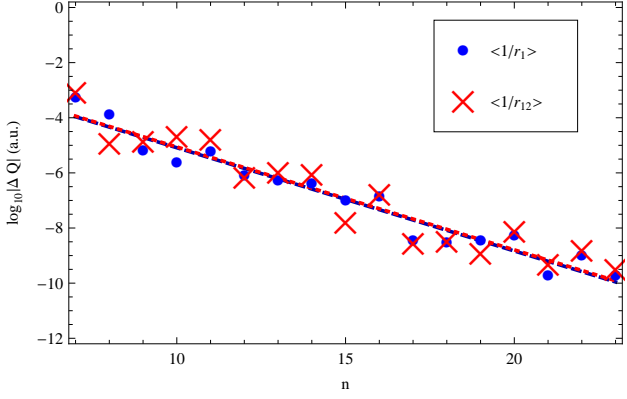}
\caption{\label{rIn} 
(Color online).
The logarithm base 10 of the error ($\Delta Q$) in the expectation values of operators that 
scale as $1/\rho$ for helium. The dark blue circles are for $\langle 1/r_1\rangle$ and
the light red crosses are for $\langle 1/r_{12} \rangle$ 
with dashed blue and dotted red fits, 
respectively (see Tab. \ref{ConvTab}).}
\end{figure}

Figure \ref{rIn} displays results for expectation values
related to the potential energy of charged particles, i.e. quantities
scaling like $1/\rho$. This probes the treatment of the singularities. The high 
degree of accuracy is evidence that these singularities have been treated correctly.

\begin{figure}[h!]
\includegraphics[width=\linewidth]{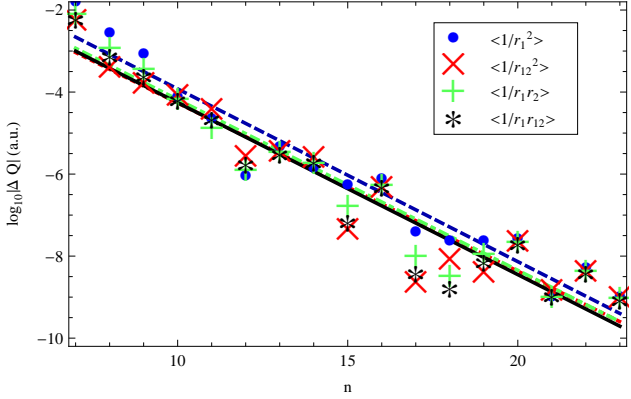}
\caption{\label{rSqIn} 
(Color online).
The logarithm base 10 of the error ($\Delta Q$) in the expectation values of operators that 
scale as $1/\rho^2$ for helium. The dark blue circles are for $\langle 1/r_1^2\rangle$,
the light red crosses are for $\langle 1/r_{12}^2 \rangle$, the green pluses are for 
$\langle 1/r_1 r_2\rangle$, and the black stars are for $\langle 1/r_1r_{12}\rangle$ with dashed blue, dotted red, dot-dashed green, and solid black fits, 
respectively (see Tab. \ref{ConvTab}).}
\end{figure}

Figure \ref{rSqIn} displays results for expectation values
related to the square of the potential energy, i.e. quantities
scaling like $1/\rho^2$. These operators emphasize the singularities even further. One
may expect that at a high enough inverse power of $\rho$ that the effect of the 
Fock logarithm become important and slow down convergence, but no evidence of that 
effect is apparent.

\begin{figure}[h!]
\includegraphics[width=\linewidth]{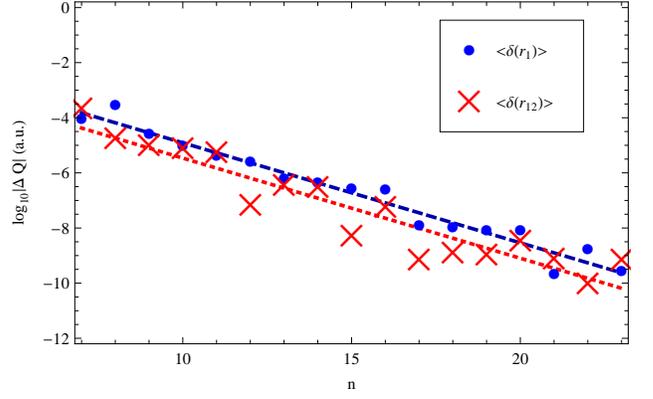}
\caption{\label{Delta} 
(Color online).
The logarithm base 10 of the error ($\Delta Q$) in the expectation values of delta function 
operators for helium. The dark blue circles are for $\langle\delta(r_1)\rangle$ and
the light red crosses are for $\langle\delta(r_{12})\rangle$ with dashed blue and dotted red fits, 
respectively (see Tab. \ref{ConvTab}).}
\end{figure}

Even the expectation values of delta functions, related to sum rule $S(2)$ (Eq. \ref{SumRulesB}), 
the Darwin term $\hat{H}_\textrm{D}$ (Eq. \ref{HDarwin}), 
and the spin-spin contact term $\hat{H}_\textrm{SSC}$ 
(Eq. \ref{HSpinSpinContact}), which are most sensitive to 
the Kato cusp conditions \cite{Kato1957} have the same convergence properties (See Fig. \ref{Delta}). 
This provides evidence that our choices of coordinates allowed the pseudospectral 
method to deduce and represent the solution in the vicinity of a cusp. It also shows that if one can handle the 
non-analyticities of the matrix element by hand, as is possible 
for delta functions (see appendix \ref{SecQuadrature}), one can still have exponentially fast convergence. 

\begin{figure}[h!]
\includegraphics[width=\linewidth]{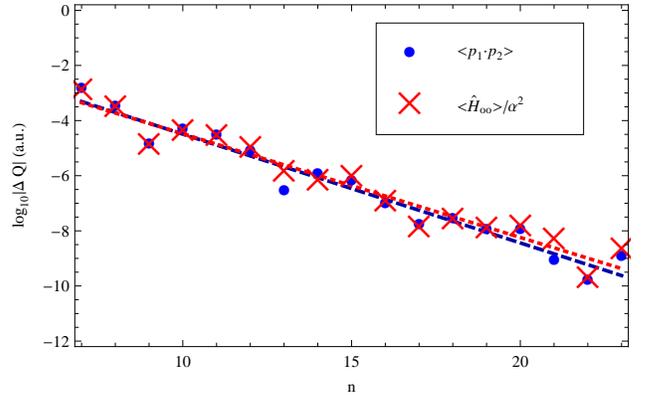}
\caption{\label{MPBP} 
(Color online).
The logarithm base 10 of the error ($\Delta Q$) in the expectation values of 
the mass polarization and the orbit-orbit interaction
operators for helium. The dark blue circles are for $\langle\bd{p}_1\cdot\bd{p}_2\rangle$ and
the light red crosses are for $\langle\hat{H}_{\textrm{OO}}\rangle/\alpha^2$ with dashed blue and dotted red fits, 
respectively (see Tab. \ref{ConvTab}).}
\end{figure}

The error in the mass polarization $\hat{H}_{\textrm{mp}}$ (Eq. \ref{HMassPolarization}),
used for the finite-nuclear mass correction and the calculation of the sum rule $S(1)$ (Eq. \ref{SumRulek1}),
and the orbit-orbit terms $\hat{H}_{\textrm{OO}}$ (Eq. \ref{HOrbitOrbit}), i.e. quadratic momentum contributions,
are shown in Fig. \ref{MPBP}. 
Calculations of derivatives (needed to form the appropriate operators) 
appear to be just as accurate as the function values, even when they are most strongly weighted 
close to the electron-electron cusp, as is the case for the orbit-orbit interaction. 

The exponential rate of convergence and the magnitude of the errors are
roughly the same in all the calculations of expectation values in
Figs. \ref{rSq}-\ref{MPBP}. This is reflected in the fits (see Tab. \ref{ConvTab}).

These errors decrease until they reach roughly the level of error
produced by truncating the wave function (see
Sec. \ref{MatrixMethods}) at the highest resolutions. 
The only easily discernible differences are at low
resolution for which the representation of the wave function at large
$\rho$ is certainly poor. It is unsurprising that the expectation
values that scale as $\rho^2$ and $\rho$ have larger errors at
low resolution due to the scarcity of points in the asymptotic tail of the 
wave function.

\begin{table}
\caption{\label{ConvTab} The fit parameters to all the convergence plots of quantities $Q$ in this section.}
\begin{ruledtabular}
\begin{tabular}{|c|c|c|c|}
$Q$ & Figure & $A$ & $\beta$\\
\hline
$\langle r_1^2\rangle$                 & \ref{rSq}  & $3.3\times 10^{-11}$ & 0.54\\
$\langle r_{12}^2\rangle$              & \ref{rSq}  & $1.2\times 10^{-10}$ & 0.53\\
$\langle \bd{r_1}\cdot\bd{r_2}\rangle$ & \ref{rSq}  & $4.9\times 10^{-11}$ & 0.47\\
$\langle r_1\rangle$                   & \ref{r}    & $1.2\times 10^{-11}$ & 0.48\\
$\langle r_{12}\rangle$                & \ref{r}    & $7.5\times 10^{-11}$ & 0.46\\
$\langle 1/r_1\rangle$                 & \ref{rIn}  & $1.1\times 10^{-10}$ & 0.37\\
$\langle 1/r_{12}\rangle$              & \ref{rIn}  & $1.2\times 10^{-10}$ & 0.37\\
$\langle 1/r_1^2\rangle$               & \ref{rSqIn}& $7.1\times 10^{-10}$ & 0.36\\
$\langle 1/r_{12}^2\rangle$            & \ref{rSqIn}& $3.1\times 10^{-10}$ & 0.38\\
$\langle 1/r_1r_2\rangle$              & \ref{rSqIn}& $3.9\times 10^{-10}$ & 0.37\\
$\langle 1/r_1r_{12}\rangle$           & \ref{rSqIn}& $2.6\times 10^{-10}$ & 0.37\\
$\langle\delta(r_1)\rangle$            & \ref{Delta}& $2.4\times 10^{-10}$ & 0.36\\
$\langle\delta(r_{12})\rangle$         & \ref{Delta}& $6.5\times 10^{-11}$ & 0.36\\
$\langle\bd{p}_1\cdot\bd{p}_2\rangle$  & \ref{MPBP} & $2.4\times 10^{-10}$ & 0.39\\
$\langle\hat{H}_{\textrm{OO}}\rangle/\alpha^2$  & \ref{MPBP} & $4.3\times 10^{-10}$ & 0.38
\end{tabular}
\end{ruledtabular}
\end{table}
All convergence data were fit to functions of the form $A\times 10^{-\beta(n-23)}$ using the same procedure as 
in Ref. \cite{GrabowskiChernoff2010}. The fit parameters are shown in Tab. \ref{ConvTab}. 
The most striking feature is how similar the magnitudes of the errors are at $n=23$. 
Also, the exponential parameter $\beta$ is roughly the same for all expectation values and
the energies and oscillator strengths (see Tab. \ref{ConvTabEOS}) with the differences already 
discussed.
Indeed, as one increases resolution
one increases the accuracy of all expectation values or oscillator strengths by roughly the same amount. 

The contributions to the total energy of the ground state of ${}^4$He are summarized in Tab. \ref{EnergyTotal}. The
values from both this work and Drake's \cite{Drake1996} are given. For a wave function with a much lower precision in 
its eigenvalue (nine decimal places compared to fifteen), nearly the same precision is obtained for the corrections to 
this eigenvalue.
 
\begin{table}
\caption{\label{EnergyTotal} The energy contributions to the ground state of ${}^4$He. These data use values of 
the physical constants $1/\alpha = 137.035999679$ and $m_e/m_\alpha = 0.000137093355571$, where $\alpha$ is the 
fine-structure constant, $m_e$ is the mass of the electron, and $m_\alpha$ is the mass of an alpha particle \cite{MohrEtAl2008a,MohrEtAl2008b}. The errors do not include the uncertainties in
these values.}
\begin{ruledtabular}
\begin{tabular}{|c|c|c|}
Energy & This Work\footnote{Values come from the $n=23$ calculation. The errors are calculated by assuming an uncertainty five times greater than the fits given in Tab. \ref{ConvTab} to account for the spread about these fits.} & Drake \cite{Drake1996}\\
\hline
$\langle \hat{H}_0 \rangle$ & -2.9037243764(8) & -2.9037243770341195 \\
$\langle \hat{H}_\textrm{mass} \rangle$\footnote{Direct evaluation of the operators $p_i^4$ ($i=1,2$) on the ket yields
delta function contributions which are unsuitable for direct numerical evaluation on the grid. So Eq. \ref{ExpValFormula} 
cannot be used to produce an exponentially accurate expectation value. As is well known, instead applying $p_i^2$ to 
both the bra and ket produces well-behaved functions, but we do not carry out this calculation in this article.} & & $-7.2006570459(3)\times 10^{-4}$ \\
$\langle \hat{H}_\textrm{OO} \rangle $ & $-7.4069807(1)\times 10^{-6}$ & $-7.40698061439(5)\times 10^{-6}$ \\
$\langle \hat{H}_\textrm{D}\rangle $ & $5.879572027(5)\times 10^{-4}$ & $5.8795720265(4)\times 10^{-4}$ \\
$\langle \hat{H}_\textrm{SSC}\rangle $ & $3.55818982(1)\times 10^{-5}$ & $3.558189840(7)\times 10^{-5}$ \\
$\langle \hat{H}_\textrm{mp}\rangle $ & $2.18103579(2)\times 10^{-5}$ & $2.1810357753732\times 10^{-5}$ \\
\end{tabular}
\end{ruledtabular}
\end{table}

\section{Conclusions}
\label{Conclusions}
We developed a general prescription for choosing coordinates and
subdomains for a pseudospectral treatment of partial differential
equations in the presence of physical and coordinate-related
singularities.  This prescription was applied to Schr\"{o}dinger's
equation for helium to determine the fully correlated
wave function. The treatment accounts for
two-body but not three-body coalescences.
Other problems with Coulomb singularities can now be tackled with
this method.

We explored the fidelity of the pseudospectral method's results.  The
method attained exponentially fast convergence for a wide selection of
expectation values and matrix elements like the oscillator strength.
Variational approaches minimize energy-weighted errors but generally do not
yield comparable results for other operators. In contrast, we found
that the pseudospectral method produced errors and convergence rates
that were very similar for all the quantities studied including energy.

The approach should be widely applicable.  No fine tuning was done to
improve convergence other than ensuring non-analytic behavior was
treated properly. The numerical
method we developed was capable of solving the large matrix problems with modest
computational resources. The calculations were pushed to the limits of
double precision arithmetic. Higher precision floating point
arithmetic will be necessary to go further.

This work generalized our previous treatment from S to P states and
demonstrated the calculation of a variety of matrix elements. It
can be further extended to higher angular momenta in a straightforward
manner, albeit at larger computational cost. 

The oscillator strength of the helium $1^1$S $\to 2^1$P transition was
calculated to about the same accuracy as the most accurate value in
the literature \cite{Drake1996} and was found to agree to the expected
precision.

\appendix

\section{Bhatia and Temkin Hamiltonian}
\label{BTAppendix}
Bhatia and Temkin \cite{BhatiaTemkin1964} derived and we checked
the following explicit expressions that make up the Hamiltonian in their
three-three splitting:
\begin{widetext}
\begin{eqnarray}
\label{BhatiaHamiltonian}
\hat{H}_S&=& -\frac{1}{2}\sum_{i=1}^2\frac{1}{r_i^2}\left(\pd{r_i}r_i^2\pd{r_i}+\frac{1}{\sin\theta_{12}}\pd{\theta_{12}}\sin\theta_{12}\pd{\theta_{12}}\right)+ \hat{V}\\
\hat{V}&=&-\frac{Z}{r_1}-\frac{Z}{r_2}+\frac{1}{r_{12}}\\
\hat{H}_{\nu,\kappa,-1}^\gamma&=& (1-\delta_{0\kappa}-\delta_{1\kappa}+(-1)^j\delta_{2\kappa})
h_\nu^\gamma B_{l\kappa,-1}
\left\{\begin{array}{ll}
\cot\theta_{12} & \textrm{ if }\nu=\gamma\\
(-1)^\nu & \textrm{ if }\nu\ne\gamma
\end{array}\right.\\
\hat{H}_{\nu\kappa 0}^\gamma&=& h_\nu^\gamma
\left\{\begin{array}{ll}
2\frac{l(l+1)-\kappa^2}{\sin\theta_{12}}+\kappa^2\sin\theta_{12}-\gamma\cot\theta_{12}l(l+1)\delta_{1\kappa}&\textrm{ if }\nu=\gamma\\
\nu\kappa(2\cos\theta_{12}+4\sin\theta_{12}\pd{\theta_{12}})-l(l+1)\delta_{1\kappa}&\textrm{ if }\nu\ne \gamma
\end{array}\right.\\
\hat{H}_{\nu\kappa1}^\gamma&=& (1-\nu \delta_{0\kappa})h_\nu^\gamma B_{l,\kappa+2,1}
\left\{\begin{array}{ll}
\cot\theta_{12} & \textrm{ if }\nu=\gamma\\
(-1)^\gamma & \textrm{ if }\nu\ne\gamma
\end{array}\right.\\ 
h_\nu^\gamma&=&\frac{1}{8\sin\theta_{12}}\left(\frac{1}{r_2^2}+\frac{\nu\gamma}{r_1^2}\right)\\
B_{l\kappa n}&=&(1+\delta_{2\kappa}(\sqrt{2}-1))^n\sqrt{(l-\kappa+1)(l-\kappa+2)(l+\kappa)(l+\kappa-1)}.
\end{eqnarray}
\end{widetext}

\section{Matrix methods}
\label{MatrixMethods}
\input{MatrixMethods}

\section{Calculating matrix elements with Bhatia and Temkin's radial functions}
\label{BTOscStrength}
\subsection{Oscillator Strength}

In the Bhatia and Temkin three-three splitting \cite{BhatiaTemkin1964},
 the matrix elements for an ${}^1$S $\to {}^1$P oscillator strength transition are written:
\begin{equation}
\sum_m|\langle {}^1\textrm{S}|\hat{\bd{D}}|{}^1\textrm{P} m \rangle|^2
=\left[\int d\tau g_{000}^0
\left(d_D^0g_{110}^0+d_D^1g_{110}^1\right)\right]^2,
\end{equation}
where $d\tau = r_1^2r_2^2\sin\theta_{12}dr_1dr_2d\theta_{12}$, $\hat{\bd{D}}$ is one of the operators found inside the matrix elements of Eqs. \ref{OscStrengthEq} and
the operators $d_D^i$ are given by
\begin{eqnarray}
d_{\bd{R}}^0&=&(r_1+r_2)\cos\frac{\theta_{12}}{2}\\
d_{\bd{R}}^1&=&(r_1-r_2)\sin\frac{\theta_{12}}{2}\\
d_{\bd{P}}^0&=&\frac{(r_1+r_2)(3+\cos\theta_{12})}{4r_1r_2\cos\frac{\theta_{12}}{2}} \nonumber\\
						& &+\cos\frac{\theta_{12}}{2} \left(\pd{r_1}+\pd{r_2}\right)\nonumber\\
						& &-\frac{(r_1+r_2)\sin\frac{\theta_{12}}{2}}{r_1r_2}\pd{\theta_{12}}\\
d_{\bd{P}}^1&=&\frac{(r_1-r_2)(-3+\cos\theta_{12})}{4r_1r_2\sin\frac{\theta_{12}}{2}}\nonumber\\
						& &+\sin\frac{\theta_{12}}{2}\left(\pd{r_1}-\pd{r_2}\right) \nonumber\\
						& &-\frac{(r_1-r_2)\cos\frac{\theta_{12}}{2}}{r_1r_2}\pd{\theta_{12}}\\
d_{\bd{A}}^0&=&\frac{Z(r_1^2+r_2^2)\cos\frac{\theta_{12}}{2}}{r_1^2r_2^2}\\
d_{\bd{A}}^1&=&\frac{Z(r_1^2-r_2^2)\sin\frac{\theta_{12}}{2}}{r_1^2r_2^2}.
\end{eqnarray}

\subsection{Expectation Values}

Similarly, an expectation value for an S state is calculated by
\begin{equation}
\label{ExpValFormula}
\langle {}^1\textrm{S}| \hat{\bd{D}} | {}^1 \textrm{S} \rangle = \int d\tau g^0_{000} d^0_D g^0_{000}.
\end{equation}
Most of the operators $d^0_D$ used for expectation values in this article have trivial forms. We write here
only the two most complicated ones:
\begin{eqnarray}
d^0_{\bd{p}_1\cdot\bd{p}_2} &=& \frac{1}{r_1 r_2}
\left[ \sin\theta_{12}\left(r_1\pd{r_1}+r_2\pd{r_2}\right)\pd{\theta_{12}} \right.\nonumber\\
& & -r_1 r_2 \cos\theta_{12}\pde{r_1}{r_2}+\cos\theta_{12}\pdd{\theta_{12}}\nonumber\\
& &\left. +\frac{1}{\sin\theta_{12}}\pd{\theta_{12}}\right] \\
d^0_{H_{OO}} &=& -\frac{\alpha^2}{2r_{12}^3}
\left[ \sin\theta_{12}\left(x_{12}\pd{r_1}+x_{21}\pd{r_2}\right)\pd{\theta_{12}} \right. \nonumber\\
& &+r_1 r_2 z_+\pde{r_1}{r_2}+z_-\pdd{\theta_{12}}\nonumber\\
& & \left. +\frac{r_{12}^2}{r_1 r_2\sin\theta_{12}}\pd{\theta_{12}}\right],
\end{eqnarray}
where 
\begin{equation}
x_{ij}=\frac{r_i^2+r_{12}^2-\bd{r}_1\cdot\bd{r}_2}{r_j}
\end{equation}
and 
\begin{equation}
z_\pm = (1\pm 3)\cos\theta_{12}(\cos\theta_{12}-\rho^2/2r_1r_2)+\sin^2\theta_{12}.
\end{equation}
All of these forms must be converted to the appropriate coordinates in each subdomain.

\input{longtabledata}

\begin{acknowledgments}

We thank Harald P. Pfeiffer for help in solving large
pseudospectral matrix problems, Saul Teukolsky and Cyrus
Umrigar for guidance and support, and Charles Schwartz for useful comments on the manuscript. 
This material is based upon work supported by the National
Science Foundation under Grant No. AST-0406635 and by NASA under Grant
No. NNG-05GF79G.

\end{acknowledgments}

\bibliography{OscStrength}

\end{document}

%% file: MatrixMethods.tex
\subsection{Formalism}

To solve for the wave function with given $k$, $l$ and
$s$ and any $m$, one must calculate the values of
$g_{\kappa ls}^\nu$ for each $\kappa$ and $\nu$ that
enters the summation in Eq. \ref{WaveFunction}. In this
section we suppress writing $k$, $l$, $s$ and $m$ indices;
only $\nu$ and $\kappa$ will appear explicitly. There
are two types of conditions which must be satisfied:
the Schr\"{o}dinger equation and the boundary
conditions.

The $\kappa$ values of interest are $\kappa_m$, the minimum value,
$\kappa_m+2$, ... up to $\kappa_M$, the maximum value. The
minimum and maximum values depend upon parity, $l$ and $\nu$ (for
notational clarity omitted).
The minimum $\kappa$ is
\begin{equation}
\kappa_m = \nu +\frac{1}{2}(1-(-1)^\nu k)
\end{equation}
and the maximum is
\begin{equation}
\kappa_M = 2 \left\lfloor \frac{l}{2} \right\rfloor - \frac{(-1)^l}{2}(1-k).
\end{equation}

Let $g_\kappa^\nu$ stand for all the grid point values for a given
$\nu$ and $\kappa$.
Assemble these in a column vector form that enumerates the
full set of $\kappa$ for a fixed $\nu$
\begin{equation}
g^\nu = \left(\begin{array}{c}
g_{\kappa_m}^\nu\\
g_{\kappa_m+2}^\nu\\
\vdots \\
g_{\kappa_M}^\nu
\end{array}\right).
\end{equation}
The length of this column vector is ${\tilde l} = 1+(\kappa_M-\kappa_m)/2$,
which takes on the values $\lfloor l/2 \rfloor$ or
$\lceil l/2 \rceil$. The size of the matrix problem increases linearly with $l$.

The Schr\"{o}dinger equation can be represented in matrix form:
\begin{equation}
\label{SchrodingerMatrixEquation}
\left(\begin{array}{cc}
H_0^0+(H_S-E)\bd{1} & H_1^0\\
H_0^1               & H_1^1+(H_S-E)\bd{1}
\end{array}\right)
\left(\begin{array}{c}
g^0\\
g^1
\end{array}\right)
=0,
\end{equation}
where $E$ is the energy, $H_S$ is the S-wave part and
$H_\nu^\gamma$ the non-S-wave part of the Hamiltonian, and $\bd{1}$ is the
identity matrix. $H_\nu^\gamma$ and $\bd{1}$ are square matrices
with dimensions ${\tilde l} \times {\tilde l}$. Explicitly, $H_\nu^\gamma$ is the tridiagonal matrix
\begin{widetext}
\begin{equation}
\label{TridiagonalHamiltonian}
H_\nu^\gamma=\left(\begin{array}{ccccc}
H_{\nu, \kappa_m, 0}^\gamma      & H_{\nu, \kappa_m, 1}^\gamma     & 0               & \cdots       & 0  \\
H_{\nu,\kappa_m+2,-1}^\gamma  & H_{\nu,\kappa_m+2,0}^\gamma & H_{\nu,\kappa_m+2,1}^\gamma   & \ddots & \vdots \\
0               & H_{\nu,\kappa_m+4,-1}^\gamma  & H_{\nu,\kappa_m+4,0}^\gamma & \ddots       & 0     \\
\vdots          & \ddots        & \ddots          & \ddots       & H_{\nu,\kappa_M-2,1}^\gamma\\
0               & \cdots        & 0               & H_{\nu,\kappa_M,-1}^\gamma & H_{\nu,\kappa_M,0}^\gamma
\end{array}\right).
\end{equation}
\end{widetext}
The third subscript on the $H^\gamma_{\nu,\kappa,n}$ labels the coupling of
the individual $g$ functions in $\kappa$.
For the S and P states calculated in this article, $H_\nu^\gamma$
is only a one by one matrix.

The pseudospectral matrices
$H_S$ and $H_{\nu,\kappa,n}^\gamma$ (for specific $\nu$, $\kappa$, $\gamma$ and $n$)
are constructed from Eq. \ref{PSMatrix} with
$\hat{H}$ replaced by $\hat{H}_S$ or $\hat{H}_{\nu\kappa n}^\gamma$, respectively (see appendix
\ref{BTAppendix} for explicit forms of these operators).
These single elements are large matrices having
dimensions set by the number of grid points.
For multiple subdomains, they are block diagonal.
The pseudospectral matrix is constructed for the
subdomain's grid points. The number of columns and
rows of an element equals the total number of grid points in all
the subdomains.

The boundary conditions can be written as
\begin{equation}
\left(\begin{array}{cc}
B_0 & 0\\
0     & B_1
\end{array}\right)
\left(\begin{array}{c}
g^0\\
g^1
\end{array}\right)
=0,
\end{equation}
where
\begin{equation}
B_\nu=\left(\begin{array}{cccc}
B_\nu^{j_m} & 0 & \cdots & 0\\
0 & B_\nu^{j_{m+2}} & \ddots & \vdots\\
\vdots & \ddots & \ddots & 0\\
0 & \cdots & 0 & B_\nu^{j_M}
\end{array}\right),
\end{equation}
is a diagonal matrix of the same size as $H_\nu^\gamma$,
and $j_m=\nu+\kappa_m+l+s$ and $j_M=\nu+\kappa_M+l+s$. Each
$B_\nu^{j}$ is a rectangular matrix of the same width as
$H_{\nu \kappa n}^\gamma$, but a smaller
height corresponding to the number
of grid points near internal boundaries or where a symmetry condition holds.
If $j$ is even (odd) $B_\nu^{j}$ enforces
zero derivative (value) along the symmetry plane.

As in Ref. \cite{GrabowskiChernoff2010}
each of the $B_\nu^{j}$ matrices can be split into two sub-matrices
\begin{equation}
\label{BCMatEq}
B_\nu^{j}=\left(B_{\nu 1}^{j} B_{\nu 2}^{j}\right),
\end{equation}
and similarly splitting the vector $g_{\kappa}^\nu$
\begin{equation}
g_{\kappa}^\nu=\left(\begin{array}{c}
g_{\kappa 1}^\nu\\
g_{\kappa 2}^\nu
\end{array}\right),
\end{equation}
yields the equation
\begin{equation}
B_{\nu 1}^{j}g_{\kappa 1}^\nu+B_{\nu 2}^{j}g_{\kappa 2}^\nu=0,
\end{equation}
where the vector and matrix have been ordered so that the index 1 refers to the
$n_b$ boundary points and the index 2 refers to the $n_i$ interior points.
The grid point nearest to the boundary, at which
an explicit boundary condition is given is considered a boundary point.
$B_{\nu 1}^{j}$ is an $n_b$ by $n_b$ matrix and $B_{\nu 2}^{j}$ is an
$n_b$ by $n_i$ matrix.
The total number of grid points is $n_t=n_b+n_i$.

Each $n_t$ by $n_t$ block of the Hamiltonian matrix $H_\nu^\gamma$
(Eq. \ref{TridiagonalHamiltonian})
can be split in a similar way,
\begin{equation}
\label{HamMatEq}
H_{\nu \kappa n}^\gamma=
\begin{array}{c} n_b \{ \\ n_i \{ \end{array}
\underbrace{\left(\begin{array}{c} H_{\nu\kappa n 11}^\gamma \\ H_{\nu\kappa n 21}^\gamma \end{array}
\begin{array}{c} H_{\nu \kappa n 12}^\gamma \\ H_{\nu\kappa n 22}^\gamma\end{array}\right)}_{n_b+n_i}.
\end{equation}

There are $n_t+n_b$ equations and $n_t$ unknowns ($g_1$ and $g_2$)
as well as the eigenvalue. One could approximately solve these equations with
singular value decomposition \cite{NumericalRecipes}, but it is much faster to
simply discard the first $n_b$ rows of each $H_{\nu\kappa n}^\gamma$ (one should still check after
finding a solution that it approximately satisfies those rows of the matrix equation)
and incorporate the boundary
conditions into the remaining eigenvalue problem by replacing each $H_{\nu\kappa n}^\gamma$ with
\begin{equation}
\label{EigenEquation}
H_{\nu\kappa n}^\gamma \to H_{\nu\kappa n 22}^\gamma-H_{\nu\kappa n 21}^\gamma (B_{\nu 1}^{j})^{-1}B_{\nu 2}^{j},
\end{equation}
where $B_{\nu 1}^{j}$ has an inverse because all of its rows are linearly independent
(otherwise more than one boundary condition would have been specified for a given
boundary point). Calculating the inverse is computationally
inexpensive since $n_b\ll n_t$. The eigenvector gives $g_{\kappa 2}^\nu$ and one
solves for $g_{\kappa 1}^\nu$ with
\begin{equation}
g_{\kappa 1}^\nu=-(B_{\nu 1}^{j})^{-1}B_{\nu 2}^{j}g_{\kappa 2}^\nu.
\end{equation}

\begin{table}
\caption{\label{MatrixData} The matrix sizes $n_m$ and number of non-zero elements $n_{NZ}$ for each resolution $n$.}
\begin{ruledtabular}
\begin{tabular}{|c|c|c|c|c|}
Resolution & \multicolumn{2}{c|}{${}^1$S States} &\multicolumn{2}{c|}{${}^1$P States}\\
\hline
$n$& $n_m$  & $n_{NZ}$    & $n_m$  & $n_{NZ}$\\
\hline
7  &  1 512 &    182 952 &  3 024 &     573 720\\
8  &  2 352 &    381 024 &  4 704 &   1 204 448\\
9  &  3 456 &    722 304 &  6 912 &   2 297 276\\
10 &  4 860 &  1 273 320 &  9 720 &   4 069 800\\
11 &  6 600 &  2 118 600 & 13 200 &   6 798 440\\
12 &  8 712 &  3 362 832 & 17 424 &  10 826 640\\
13 & 11 232 &  5 133 024 & 22 464 &  16 571 568\\
14 & 14 196 &  7 580 664 & 28 392 &  24 531 416\\
15 & 17 640 & 10 883 880 & 35 280 &  35 292 600\\
16 & 21 600 & 15 249 600 & 43 200 &  49 536 960\\
17 & 26 112 & 20 915 712 & 52 224 &  68 048 960\\
18 & 31 212 & 28 153 224 & 62 424 &  91 722 888\\
19 & 36 936 & 37 268 424 & 73 872 & 121 570 056\\
20 & 43 320 & 48 605 040 & 86 640 & 158 726 000\\
21 & 50 400 & 62 546 400 &        &\\
22 & 58 212 & 79 517 592 &        &\\
23 & 66 792 & 99 987 624 &        &
\end{tabular}
\end{ruledtabular}
\end{table}

\subsection{Matrix Eigenvalue Solution}
\label{MatEigSol}

The number of grid points in each sub-domain, $\{x,\phi,\theta_{12}\}$
or $\{x,\zeta,\beta_{12}\}$,
was $n_t=2n\times n\times n$; greater resolution is needed along
the semi-infinite coordinate. This leads to a Hamiltonian matrix size of $n_t\times n_t$
for S states and $2n_t \times 2 n_t$ for odd parity P states.
After solving for boundary conditions with the above procedure, these are reduced to $n_m\times n_m$
and $2n_m\times 2n_m$, respectively,
where $n_m=n_i=6n^3-12n^2+6n$.
The number of non-zero elements $n_{NZ}$ scales as $n^4$. For
$n=20$, this corresponds to 560 MB and 1.8 GB, respectively, of memory required to
store the matrix.\footnote{Note: some eigenvalue solvers do not require one to store this matrix and simply
require a function which can calculate the matrix times a given vector.} The sizes of the matrices and the
number of non-zero elements is given in Tab. \ref{MatrixData}.

The method of inverse iteration \cite{NumericalRecipes} was used to find eigenvalues
with a shift equal
to the known eigenvalues plus $10^{-4}$ so that the matrix is not too singular. Each iteration
requires a matrix solve. For the smaller matrices (up to $17,000\times 17,000$),
these solves were performed using
Mathematica's \cite{Mathematica2008} multifrontal matrix solve routine. This method is fast (eigenvalues can
be calculated in about 10 minutes for that size) but $8$ GB of RAM was insufficient to do
larger sizes. For larger matrices,
the {\it g}eneralized {\it m}inimal {\it res}idual (GMRES) method of PETSc
\cite{petsc-web-page,petsc-user-ref,petsc-efficient} was used. The GMRES method produces a solution
with the Krylov space of the matrix and is more memory efficient.

\begin{figure}
\includegraphics[width=\linewidth]{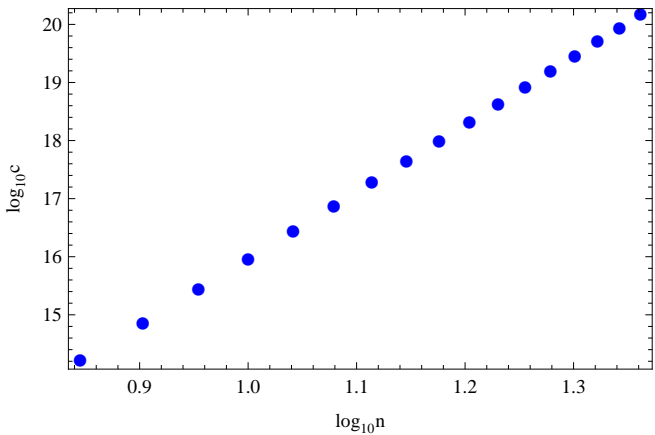}%
\caption{\label{CondNum}
(Color online).
A log-log plot of the spectral condition number $c$ of the pseudospectral matrices
as a function of resolution.
}
\end{figure}

Preconditioning is essential for solving large matrix problems. A measure of how hard a matrix problem
is to solve (how fast a method converges) is the spectral condition number, defined as
\begin{equation}
c = \frac{|\lambda_\textrm{max}|}{|\lambda_\textrm{min}|},
\end{equation}
where $\lambda_\textrm{max}$ and $\lambda_\textrm{min}$ are the eigenvalues with the largest and smallest magnitudes,
respectively.
The spectral condition numbers of pseudospectral matrices grow rather fast with increasing resolution
\cite{PfeifferCom,Fornberg1996sec54}.
For the problem at hand, it is plotted versus resolution in Fig. \ref{CondNum}. It starts
out large and grows asymptotically as $n^{12}$. An ill-posed problem has a condition number which
grows exponentially \cite{TeukolskyCom}. This problem is well-posed but in order to solve this system of equations
preconditioning is necessary.
A reasonable preconditioner is a matrix produced by a second order finite
differencing scheme on the same set of grid points \cite{PfeifferCom,Orszag1980,Fornberg1996sec54}.
The preconditioning matrix
solves are further preconditioned with a block Jacobi preconditioner.

The modified Gramm-Schmidt procedure was used to orthogonalize the Krylov subspace. Furthermore, the
GMRES restart parameter, $m$, needs to be very large for convergence, empirically, $m=1.3 n_m^{3/4}$,
where $n_m\times n_m$ is the matrix size. The computation time scales as $n_m^3$, which for 
the largest matrix size was about a day running on six 2 GHz processors.
The eigenvalue solver is the slowest part of the entire computation.

All calculations were done with double precision arithmetic. This gives some minimum
error in the calculated eigenstate. The effect is relatively big for the small exponential tail. The key observation is that
the wave function no longer decreases at the theoretically
expected asymptotic rate when it drops to about
$10^{-8.7}$ of its maximum value, after which it takes on a seemingly random value less than this magnitude.
This value is independent of resolution because of the limits of machine precision arithmetic.
It is possible that the asymptotic tail could be better calculated with a better preconditioner.

The issue of the asymptotic behavior is important.
Since a constant value
for the wave function on a semi-infinite domain leads to divergent matrix elements,\footnote{For a 
finite resolution, the quadrature still leads to a finite result with an error
enhanced by at most $10^4$ for the cases calculated in this article.} we set any value of the eigenvector below this threshold to zero.

\subsection{Quadrature}
\label{SecQuadrature}

In this article, it is
necessary to calculate matrix elements of the form $\langle i|\hat{O}|j\rangle$, where $|i\rangle$
and $|j\rangle$ are two quantum states and $\hat{O}$ is some operator. This calculation requires
numerical integration. Pseudospectral methods, by design, use quadrature points as the grid points.
A one dimensional function $f[\gencoord]$ can be numerically integrated from $\gencoord=-1$ to $\gencoord=1$ with weight
function $g[\gencoord]$ by
\begin{equation}
\label{OneDQuadrature}
\int_{-1}^1f[\gencoord]g[\gencoord]d\gencoord\approx \sum w_if[\gencoord^i],
\end{equation}
where $w_i$ is the quadrature weight specific to the weighting function $g$ at grid point $\gencoord^i$.
This quadrature formula is exponentially accurate with increasing resolution if $f$ is smooth
over the domain $-1\le \gencoord\le 1$.
The problems solved in this article are three-dimensional with three overlapping subdomains.
A separate quadrature can be done in each sub-domain. This is illustrated for domain ${D}_1$ with coordinates $\{x, \phi, \theta_{12}\}$
and ranges $-1\le x \le 1$, $0\le \phi \le 1/2$, and $0\le \theta_{12}\le \pi$. Define
\begin{eqnarray}
\gencoord_1&=&x\\
\gencoord_2&=&4\phi-1\\
\gencoord_3&=&\frac{2\theta_{12}}{\pi}-1,
\end{eqnarray}
so that $-1\le \gencoord_1,\gencoord_2,\gencoord_3\le 1$.
Integrals over ${D}_1$ use three-dimensional sums analogous to Eq. \ref{OneDQuadrature}. Since the ranges are fixed, the order of nesting is immaterial.
To satisfy the requirement that $f$ is smooth (up to the logarithmic singularity at $\rho=0$),
choose $g=1$,\footnote{For each integral, one has an integrand, $\psi \hat{O} \psi$ times the factor from the
volume element. This whole product is $f[X]g[X]$ so there is freedom as to how one divides the integrand between
$f$ and $g$ up to the restriction that $f$ be smooth. The simplest choice is made here.}
which corresponds to Legendre quadrature points, which are used for all
calculations in this article instead of Chebyshev which were used in Ref.
\cite{GrabowskiChernoff2010}.

\begin{figure}
\includegraphics[width=\linewidth]{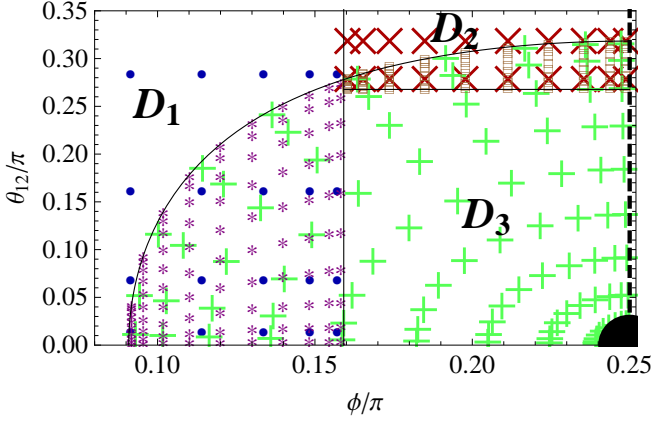}%
\caption{\label{OverlapPoints}
(Color online).
This is the arrangement of grid points of the three domains at a
constant value of $\rho$ in $\phi$ and $\theta_{12}$ coordinates for $n=10$.
As in Fig. \ref{GridPoints}, the blue circles,
red crosses, and green pluses belong to domains $D_1$, $D_2$, and $D_3$,
respectively. Also shown are the overlap grid points in $D_1\cap D_3$ (purple stars)
and $D_2\cap D_3$ (brown squares).
The electron-electron singularity is visible at the lower right hand
corner (solid disk at $\phi=\pi/4, \theta_{12}=0$) as well as the line of symmetry on
the right side (dashed line at $\phi=\pi/4$ where $r_1=r_2$).}
\end{figure}

If all the subdomains are non-overlapping, then the above scheme is sufficient for all integrals.
However, no set of non-overlapping subdomains for which $f$ is smooth could be found.\footnote{Some
do exist which are only non-analytic on some edges,
but these produce noticeable non-exponential convergence.}
A method is needed for handling overlapping regions, which the above scheme double counts if a
quadrature is performed in each sub-domain. For these
regions, an interpolation was performed to two new $2n\times n\times n$ grids spanning the
overlap regions, shown in Fig \ref{OverlapPoints}. For the pseudospectral method,
interpolation is done to the same order as the grid size. A quadrature can then be done over the
overlap regions, which are used to correct the overall integration.

The overlap region is divided into two subdomains
\begin{eqnarray}
{D}_{13}&=&{D}_1\cap {D}_3\\
{D}_{23}&=&{D}_2\cap {D}_3 .
\end{eqnarray}
These subdomains satisfy
\begin{widetext}
\begin{equation}
\begin{array}{lccc}
D_{13}:& -1\le x\le 1, & \phi_{\textrm{min}}\le \phi \le \frac{1}{2}, & 0\le \theta_{12}\le \theta_{12,\textrm{max}}[\phi]\\
D_{23}:& -1\le x\le 1, & \frac{1}{2} \le \phi \le \frac{\pi}{4}, & \arccos\frac{2}{3}\le \theta_{12}\le \theta_{12,\textrm{max}}[\phi],
\end{array}
\end{equation}
\end{widetext}
where $\phi_{\textrm{min}}=\pi/4-1/2$ is determined by $\zeta=1/2$ and $\theta_{12}=0$
and $\theta_{12,\textrm{max}}[\phi]=\arccos[\cos 1\csc 2\phi]$ is determined by $\zeta=1/2$.
One defines appropriate $\{\gencoord_1, \gencoord_2, \gencoord_3\}$. For example, in $D_{13}$
\begin{eqnarray}
\gencoord_1&=&x\\
\gencoord_2&=&2\left(\frac{\phi-\phi_{\textrm{min}}}{\frac{1}{2}-\phi_{\textrm{min}}}\right)-1\\
\gencoord_3&=&2\theta_{12,\textrm{max}}[\phi]\theta_{12}-1.
\end{eqnarray}
Now one calculates the nested sum with $\gencoord_3$ innermost since
the range of $\theta_{12}$ depends upon $\phi$.

The function values at the points necessary for the quadrature $\{x^{j_1'},\phi^{j_2'},\theta_{12}^{j_2'j_3'}\}$
are calculated with interpolation
\begin{equation}
\label{FuncInterpolation}
f[x^{j_1'},\phi^{j_2'},\theta_{12}^{j_2'j_3'}]\approx\sum_{J}f[x^{j_1},\phi^{j_2},\theta_{12}^{j_3}]
{\cal C}_{J}[x^{j_1'},\phi^{j_2'},\theta_{12}^{j_2'j_3'}].
\end{equation}
where ${\cal C}_J$ refers to the effective basis defined in
Eq. \ref{EffectiveBasis} and $J=\{j_1,j_2,j_3\}$.

Sometimes $f$ involves a Dirac delta function. In such a case, one integrates out the
delta function analytically. One is left with a two dimensional integral on the surface where the
argument of the delta function is zero. This entails first interpolating to that surface using
Eq. \ref{FuncInterpolation}. One can then proceed normally with a two-dimensional quadrature.

%% file: longtabledata.tex
\section{History of Oscillator Calculations}
\label{OscStrengthTable}

Table \ref{History1S2P-OscStrengths} summarizes the last half
century's theoretical studies of the nonrelativistic,
electric dipole oscillator strength. The prime
criterion for inclusion in the Table is that a numerical value for
the oscillator strength for the specific transition
$1^1$S $\to 2^1$P be calculated and quoted.  We do not
indicate in this Table other transitions calculated
even though these often constitute the bulk of a
paper's research results. In broadest terms, the
entries illustrate progress in achieving higher
accuracy for the specific transition and/or testing
new methods designed to yield more extensive sets of
bound-bound oscillator strengths.

Many methods appearing in Table \ref{History1S2P-OscStrengths} are variational and
utilize the exact interaction potential of the
nonrelativistic Hamiltonian
\cite{SchiffPekeris1964,GreenEtAl1966,Weiss1967,Chong1968,SchiffEtAl1971,KonoHattori1984,CannThakkar1992,Chen1994,Chen1994b,Yang1997,Drake1996}.
Variational methods are especially useful when electron
correlation is important and ground state properties are
sought.  There are many strategies for selecting
bases and suitable variational parameters. 
This flexibility may become cumbersome for the study of
highly excited states if lower level states must be
projected out as a preliminary step (e.g. if the trial
wave function is not linear in the unknown parameters
and one seeks to enforce orthogonality of the excited
state with respect to lower states). Errors in the
eigenproblem accumulate and higher levels are harder to
find accurately, even when the wave function is linear in 
the variational parameters.

A general conclusion is that
some basis choices do a better job representing
the parts of the wave function critical to oscillator
strength calculations.  Configuration interaction (CI)
calculations
\cite{SchiffPekeris1964,Brown1969,BrownCortez1972,DavisChung1982,Chen1994}
converge but suffer from the absence of odd powers of
the inter-electronic distance \cite{Drake1999}. 
Perimetric \cite{Pekeris1958} coordinates
\cite{SchiffEtAl1971,Drake1996,Drake1999,DrakeEtAl2002,Drake2004,DrakeMorton2007,DrakeYan2008}
and Hylleraas \cite{Hylleraas1929} coordinates
\cite{Weiss1967,AndersonWeinhold1974,KonoHattori1984,Yang1997}
include terms of this sort.
Systematic variational studies using bases incorporating the
inter-electronic distance have yielded some of the more accurate
calculations to date. A Hylleraas expansion is used by Drake who
determined the oscillator strengths to seven decimal
digits \cite{Drake1996}, the most precise calculations
thus far, as well as some finite-nuclear-mass and
relativistic corrections. At this stage further nonrelativistic
calculations of the oscillator strength are probably less important than the
inclusion of spin-orbit, mass polarization and low-order
relativistic effects.

Expansions in terms of orthogonal functions
often produce basis elements of
increasing complexity. Alternatively, one can
use larger numbers of simpler functions.
One important example is the exponential
basis \cite{ThakkarSmith1977,FrolovSmith1995} (exponential functions of
$r_1$, $r_2$ and $r_{12}$), which has the great advantage of 
having an easy to calculate Hamiltonian matrix at the expense of violating cusp
conditions.  This basis was used by Cann and Thakkar \cite{CannThakkar1992}
to get many different oscillator strengths for S $\to$ P and P $\to$ D transitions
of helium-like atoms. They got the $1^1$S $\to 2^1$P oscillator strength correct to
five decimal places.

The central field approximation \cite{BetheSalpeter1957} is suitable when
electrons are nearly uncorrelated and exchange effects
are negligible. The essence of this approximation is
twofold: (1) the multi-electron wave function is
written in terms of products of one-electron functions
and (2) each electron experiences a potential which is
a function only of its distance to the nucleus. The
omission of explicit inter-electronic coordinates
hinders convergence but greatly simplifies the
variational problem.  Green {\it et al.}
\cite{GreenEtAl1966} produced tables of S $\to$ P and
P $\to$ S transitions using the configuration interaction
form for the wave functions.

There exist many different approximations to
representing the fully correlated wavefunctions.
Multiconfiguration Hartree-Fock recovers some but not
all of the electron correlation energy and yields
improved oscillator strengths compared to Hartree-Fock
treatments \cite{Fischer1974}. The coupled cluster
expansion (roughly analogous to a truncated form of
configuration interaction) also yields better results
\cite{FernleyEtAl1987}.

Simplifications are frequently made to generate
comprehensive but approximate oscillator strength
databases. With this approach the physical as opposed to
numerical errors may be difficult to gauge.
For a two-electron atom the Hamiltonian may be written
\begin{eqnarray}
\hat{H}_0&\approx& \hat{H}_1+\hat{H}_2\\
\hat{H}_i&=&\frac{p_i^2}{2}+U_i[r_i],
\end{eqnarray}
where $U_i$ accounts for the screening of the nucleus
by the electron cloud.  If the matrix element is
dominated by the wave function at large distances one
may adopt the asymptotic form of the potential in that
limit to give the Coulomb approximation
\cite{BatesDamgaard1949},
\begin{equation}
U_i[r_i]\approx -\frac{Z-1}{r_i}.
\end{equation}
In this approximation
the regularity condition at $r=0$ no longer applies; one needs
an alternate method of determining the discrete energy
eigenvalues. These may be borrowed from experimental measurements or
other theoretical calculations and are referred to as ``hybrid''
results in Table \ref{History1S2P-OscStrengths}. Wiese {\it et al.}
\cite{WieseEtAl1966,WieseEtAl1969} used this approximation (with
exchange effects) to calculate oscillator strengths for the elements from hydrogen to
calcium, Cameron {\it et al.} \cite{CameronEtAl1970} tabulated 95
different transitions, and Theodosiou \cite{Theodosiou1987} produced
extensive tables with errors better than 10\% based on a more
sophisticated form \cite{Desclaux1970} of $U_i$. He calculated the
oscillator strength of the $1^1$S $\to 2^1$P transition to four
decimal places.  Runge and Valance \cite{RungeValance1983} developed a
similar approach based on the atomic Fues potential for the valence
electron,
\begin{equation}
U[r]=-\frac{Z}{r}+\sum_{l=0}^\infty \frac{B_l \hat{P}_l}{r^2},
\end{equation}
where $B_l$ is an adjustable parameter and $\hat{P}_l$
is the projection operator onto a subspace of given
angular momentum $l$.  Currently, the most complete
tabulation of transitions is given by Wiese and Fuhr
\cite{WieseFuhr2009}.

The Table includes calculations based on
perturbation theory.  Sanders, Scherr, and Knight
\cite{SandersScherr1969,SandersKnight1989} developed a
$1/Z$ expansion, in which the electron-electron
interaction is the perturbation.  Even for $Z=2$,
calculations could be carried out to high enough order
that the oscillator strengths converged to three
decimal places for the helium $1^1$S $\to 2^1$P
transition. One merit of this approach is that it
yields oscillator strength as a function of $Z$ and
with an improving accuracy as $Z$ and/or excitation
levels increases.

Devine and Stewart \cite{DevineStewart1972a,DevineStewart1972b}
divided the Hamiltonian into two parts
\begin{equation}
\hat{H}_0=\hat{H}_{HF}+\hat{H}_1,
\end{equation}
where $\hat{H}_{HF}$ is the Hamiltonian projected into
the subspace spanned by solutions of the Hartree-Fock
type and $\hat{H}_1$ is the difference between this
operator and the full nonrelativistic Hamiltonian
$H_0$.  The operator $\hat{H}_1$ was treated as a
perturbation parameter using wave functions derived from
the frozen Hartree-Fock core. They derived oscillator
strengths correct to three decimal places using
second-order perturbation theory.


Finally, some results do not attempt to calculate oscillator strengths
with greater precision or for larger sets of transitions. 
Anderson and Weinhold
\cite{AndersonWeinhold1974} calculated oscillator strengths
{\it and} rigorous bounds on those values.


The $1^1$S $\to 2^1$P oscillator strength results
derived in this paper by the pseudospectral method are not
listed in Table \ref{History1S2P-OscStrengths} but
match the accuracy of the most accurate included. 
The method has not yet been tested on transitions
involving other states.

\begin{widetext}
\begin{longtable}{|p{2.5cm}|p{5cm}|p{2.5cm}|p{5cm}|}
\caption{A brief history of theoretical 
calculations of the non-relativistic, electric dipole contribution 
to the oscillator strength for $1^1$S $\to 2^1$P transition.}\label{History1S2P-OscStrengths} \\
\hline \multicolumn{1}{|c|}{Authors} & \multicolumn{1}{c|}{Method} & \multicolumn{1}{c|}{Value} & \multicolumn{1}{c|}{Notes} \\ \hline
\endfirsthead

\multicolumn{4}{c}%
{{\bfseries \tablename\ \thetable{} -- continued from previous page}} \\
\hline \multicolumn{1}{|c|}{Authors} & \multicolumn{1}{c|}{Method} & \multicolumn{1}{c|}{Value} & \multicolumn{1}{c|}{Notes} \\ \hline
\endhead

\hline \multicolumn{4}{|r|}{{Continued on next page}} \\ \hline
\endfoot

\hline \hline
\endlastfoot

Trefftz {\it et al.} \cite{Trefftz1957} &
HF, explicit corr & 
$0.3113$L $0.2719$V&
Table 4, wf: 2 orbitals with $r_{12}$, v preferred
 \\
Dalgarno and Lynn 1957 \cite{DalgarnoLynn1957} & Sum rules & $0.239$ & Table 1, $f$'s from
earlier calculations modified for conformity\\
Dalgarno and Stewart 1960 \cite{DalgarnoStewart1960} & var, Hyll & $0.275$ & quoted Low and Stewart in
Table 2, wfs: 6 parameter S, $Z^*$ hydrogenic P \\
Schiff {\it et al.} \cite{SchiffPekeris1964} &
var, peri coord &
$0.276159$L $0.276164$V $0.276149$A &
Table I, extrap 56, 120, 220 term wfs, method D
\\
 & & 
$0.276154$L $0.276150$V &
Table VII, extrap 56, 120, 220 term wfs, method C
\\
 & &
$0.27616$ &
Table IX, $\pm 0.00001$, summary.
 \\
Green {\it et al.} \cite{GreenEtAl1966} &
var, CF with exch and CI &
$0.27537$L $0.27586$V $0.26908$A &
Table 1,
Slater orbitals, wf: 50 terms 1S, 42 terms 2P,
$Z^*$, hybrid
 \\
Weiss \cite{Weiss1967} &
var, Hyll coords &
$0.2759$L $0.2761$V &
Table 2, wf: 53 terms 1S, 52 terms 2P; EFs; hybrid
\\
Cohen and Kelly \cite{Cohen1967} &
HF, FC; one valence electron;
some exch &
$0.112$L &
Table V
\\
Dalgarno and Parkinson \cite{Dalgarno1967} &
HF, $\sim Z^{-1}$ &
$0.373$ &
Table 3, first order in $Z^{-1}$
 \\
Chong and Benston \cite{Chong1968} &
var, constrained by
off-diagonal hypervirial theorem & 
$0.26385$  &
f from M ($0.41620$, Table II) and calculated
energy ($0.77459$), wf: 7 terms for 1S, 2 terms for 2P, $Z^*$
 \\
Sanders and Scherr \cite{Sanders1969} &
var, Hyll coord, $Z^{-1}$ &
$0.276113$L $0.276182$V $0.276012$A &
Table XVIII, wfs: 100 terms, 9-th order in Z
 \\
Cameron {\it et al.} \cite{CameronEtAl1970} &
HF FC, one valence electron& $0.281$L $0.255$V
 & Table I
 \\
Schiff {\it et al.} \cite{SchiffEtAl1971} &
var, peri coord  &
$0.276165$V & Table XIV,
wfs: up to 1078 terms for S state, 364 for P states;
converged to within number of digits quoted
 \\
Devine and Stewart \cite{DevineStewart1972b} & 
HF, FC, Pert &
$0.2760$L $0.2749$V $0.2771$A &
Table 2, iterated result, wf: 77 terms for S state, 65 for P state
\\
Laughlin \cite{Laughlin1973} &
$Z^{-1}$, mod screening &
$0.29834$ &
Table 4, f from expansion coefficients
\\
Anderson and Weinhold \cite{AndersonWeinhold1974} &
rigorous limits &
$0.2747-0.2775$ &
Table IV
 \\
Froese Fischer \cite{Fischer1974} &
MCHF &
$0.2753$L $0.2744$V &
Table 2
\\
Leopold and Cohen \cite{Leopold1975} &
upper bounds &
$<0.29678$ &
bound from $\sigma^2$ (Table 1) and best NR energy; hybrid
\\
Davis and Chung \cite{DavisChung1982} &
CI, no r12 corr, AMPW&
$0.2721$L $0.2758$V &
Table V, 110 terms S and P states
\\
Roginsky and Klapisch \cite{RoginskyEtAl1983} &
Modified wf &
$0.256$L+V &
Table 1, product wfs with $Z^*$
\\
Kono and Hattori \cite{KonoHattori1984} &
var, double Hyll, ECFs &
$0.27616$ &
Table III, 138 terms S and 140 terms P, 3 nonlinear parameters (2 set, 1 optimized), $\pm 0.00001$
\\
Theodosiou \cite{Theodosiou1984} &
Valence electron in potential &
$0.2761$L &
Table I, HF Slater potential, hybrid (experimental)
\\
Park {\it et al} \cite{ParkEtAl1986} &
HSA & $0.291$L $0.342$A & Table I, initial (final) wf 4 (6) 
angular momentum pairs
\\
Theodosiou \cite{Theodosiou1987} &
as above &
$0.27643$L &
Table I
\\
Fernley {\it et al.} \cite{FernleyEtAl1987} &
CC expansion &
$0.2811$ & Table 3, 1s, 2s, 2p, {$\bar 1$}d, {$\bar 3$}p one-electron states and 
product states; R-matrix inner region, numerical integration outer region
\\
Sanders and Knight \cite{SandersKnight1989} &
var, Hyll, $\sim Z^{-1}$, pert &
$0.27774$ &
Table V, wfs and energies from \cite{Sanders1969} 
\\
Abrashkevich {\it et al.} \cite{AbrashkevichEtAl1991}
 & HSAnacc & 
$0.2763$L $0.2844$A & Table 2, initial (final) 6 (4) radial equations, 100
finite elements
\\
Cann and Thakkar \cite{CannThakkar1992} &
var, exp, ECFs &
$0.27617$ &
Table V, 100 terms, 6 nonlinear parameters
(error of $0.7-2.99$ units in last digit)
\\
Tang {\it et al.} \cite{TangEtAl1992}
 & HSCC CC & $0.2762$L $0.2763$A & Table I
\\
Chen \cite{Chen1994} &
CI with B-splines &
$0.27611$ &
Table 12, 150 9-th and 10-th order splines for S,
137 for P, uncertainty $\le 0.01$\%
\\
Chen \cite{Chen1994b} &
CI with B-splines &
$0.276163$L $0.276076$V &
Table 13, 150 9-th and 10-th order splines for S, 
147 for P, hybrid (best NR energies)
\\
Yang \cite{Yang1997} &
MELL, peri &
$0.276165$L $0.276165$V &
Table 3.7, 680 terms, 2 nonlinear parameters
\\
Drake \cite{Drake1996} &
var, double Hyll &
$0.2761647$ &
Table 11.11, nonlinear scale parameters 
\\
Masili {\it et al} \cite{MasiliEtAl2000} &
HSAnacc & $0.2761957$ & Table 4, initial (final) 13 (15) radial equations
\\
Alexander and Coldwell \cite{AlexanderColdwell2006} &
varMC &$0.2761$L $0.2706$V $0.2758$A & Table V, largest wfs,
rotated method

\end{longtable}

\begin{longtable}{|p{3.5cm}|p{10cm}|}
\caption{Abbreviations in above table.}\label{Abbreviations} \\
\hline \multicolumn{1}{|c|}{Symbol} & \multicolumn{1}{c|}{Meaning} \\ \hline
\endfirsthead
\multicolumn{2}{c}%
{{\bfseries \tablename\ \thetable{} -- continued from previous page}} \\
\hline \multicolumn{1}{|c|}{Symbol} & \multicolumn{1}{c|}{Meaning} \\ \hline
\endhead
\hline \multicolumn{2}{|r|}{{Continued on next page}} \\ \hline
\endfoot
\hline \hline
\endlastfoot
A & acceleration form for oscillator strength \\
AMPW scale params & angular momentum partial wave scale parameters \\
CC expansion & close coupling expansion \\
CF & central field (no separation coordinate) \\
CI & configuration interaction \\
corr & correlation factors \\
double Hyll & double Hylleraas coordinate basis \\
ECFs & multiple exponential correlation factors \\
EFs & multiple exponential factors \\
exch & exchange interactions \\
extrap & extrapolation based on \\
exp & exponential basis (exponentials of Hyllerass coordinates) \\
f & oscillator strength \\
FC & frozen core \\
HF & Hartree Fock \\
HSA & adiabatic Hyperspherical coordinate representation \\
HSAnacc & HSA with non-adiabatic channel coupling \\
HSCC & Hyperspherical coordinate representation; CC expansion \\
hybrid & energy not taken from parmeterized 
wave function; input from experiment or other calculations \\
Hyll coord & Hylleraas coordinates \\
L & length form for oscillator strength \\
mod screening & modified screening approximation \\
M & dipole moment \\
MCHF & multiconfiguration Hartree Fock \\
MELL & matrix expansion in exponentials, Laguerre polynomials and eigenfunctions of total orbital angular momentum \\
NR & non-relativistic \\
peri coord & perimetric coordinates \\
Pert & perturbation theory corrections \\
var & variational \\
varMC & variational Monte Carlo\\
V & velocity form for oscillator strength \\
wf, wfs & wave function, wave functions \\
$Z^{-1}$ & expansion in inverse powers of Z \\
$\sim Z^{-1}$ & expansion in inverse powers of Z with additional corrections \\
$Z^*$ & nonlinear, effective nuclear charge parameter\\

\end{longtable}
\end{widetext}